\documentclass[12pt]{article}
\usepackage{amsmath}
\usepackage{graphicx}
\usepackage{enumerate}
\usepackage{natbib}
\usepackage{amsmath}
\usepackage{url} 

\newcommand{\blind}{0}

\DeclareMathAlphabet{\mybm}{OT1}{ptm}{b}{it}

\usepackage{mathptmx}
\usepackage{euscript}
\usepackage{latexsym}
\usepackage{amsmath}
\usepackage{amsfonts}
\usepackage{amssymb}
\usepackage{ulem}
\usepackage{amsthm}

\newtheorem{thm}{Theorem}

\newtheorem{pro}{Proposition}

{\theoremstyle{remark} 
\newtheorem{rem}{Remark}}

{\theoremstyle{definition} 

}

\newcommand{\BIBO}{{\small \BIBO}}

\newcommand{\plim}{\stackrel{\raisebox{-0.06in}{$\scriptscriptstyle P$\,}}{\rightarrow}}

\newcommand{\linfty}{\rightarrow \infty}
\newcommand{\lzero}{\rightarrow 0}

\newcommand{\eq}{\begin{eqnarray*}}
\newcommand{\eqq}{\end{eqnarray*}}
\newcommand{\eqn}{\begin{eqnarray}}
\newcommand{\eqqn}{\end{eqnarray}}

\newcommand{\eqb}{\begin{align*}}
\newcommand{\eqqb}{\end{align*}}

\newcommand{\E}{\mathrm{E}}

\newcommand{\Cov}{\mathrm{Cov}}

\newcommand{\tr}{\mathrm{tr}}

\newcommand{\eqna}{\begin{align}}
\newcommand{\eqqna}{\end{align}}

\newcommand{\bA}{\mathbf{A}}

\newcommand{\bH}{\mathbf{H}}
\newcommand{\bI}{\mathbf{I}}

\newcommand{\bQ}{\mathbf{Q}}
\newcommand{\bR}{\mathbf{R}}
\newcommand{\bS}{\mathbf{S}}

\newcommand{\bU}{\mathbf{U}}
\newcommand{\bV}{\mathbf{V}}

\newcommand{\bX}{\mathbf{X}}
\newcommand{\ba}{\mathbf{a}}

\newcommand{\by}{\mathbf{y}}
\newcommand{\bz}{\mathbf{z}}

\newcommand{\bu}{\mathbf{u}}

\newcommand{\bbR}{\mathbb{R}}

\newcommand{\cF}{\EuScript{F}}

\newcommand{\al}{\alpha}

\newcommand{\ep}{\epsilon}
\newcommand{\gam}{\gamma}

\newcommand{\lam}{\lambda}

\newcommand{\om}{\omega}

\newcommand{\sig}{\sigma}

\newcommand{\del}{\delta}

\newcommand{\vsig}{\varsigma}

\newcommand{\bmgam}{\pmb{\gamma}}

\newcommand{\bmphi}{\pmb{\phi}}

\newcommand{\bmth}{{\pmb{\theta}}}

\newcommand{\bmxi}{\pmb{\xi}}

\newcommand{\bGam}{\pmb{\Gamma}}

\newcommand{\bPhi}{\pmb{\Phi}}

\newcommand{\albar}{\overline{\al}}
\newcommand{\alu}{\underline{\al}}

\usepackage{float}
\usepackage{caption}
\usepackage{subcaption}
\usepackage{booktabs}

\def\spacingset#1{\renewcommand{\baselinestretch}%
{#1}\small\normalsize} \spacingset{1}

 \DeclareMathSymbol{,}{\mathpunct}{operators}{"2C}

\begin{document}

\if0\blind
{
  \title{\bf  Quantile-Crossing Spectrum and  Spline Autoregression Estimation}
  \author{Ta-Hsin Li\footnote{Formerly affiliated with IBM T. J. Watson Research Center, Yorktown Heights, NY 10598, USA.
  Email: {\sc thl024@outlook.com}} }
  \date{September 3, 2025}
  \maketitle
} \fi

\if1\blind
{
  \title{\bf  Quantile-Crossing Spectrum and  Spline Autoregression Estimation} 
  \maketitle
} \fi

\begin{abstract}
The quantile-crossing spectrum is the spectrum of quantile-crossing processes created from a time series
by the indicator function that shows whether or not the time series lies above or below 
a given quantile  at a given time. This bivariate function of frequency and quantile level provides 
a richer view of serial dependence than that offered by the ordinary spectrum.
A new estimator is proposed in this paper for the quantile-crossing spectrum as a bivariate function of frequency and quantile level. The proposed estimator is derived from a method called spline autoregression (SAR).
It  jointly fits an autoregressive (AR) model 
to the quantile-crossing series across multiple quantiles, where the functional AR coefficients are represented as spline functions of the quantile level and penalized for their roughness. Numerical experiments show that when the underlying spectrum is smooth in quantile level the proposed method is able to produce more accurate estimates in comparison with the alternative that ignores the smoothness.

\vspace{1in}
\noindent
{\it Keywords}: autoregression, penalized least squares,  quantile-frequency analysis, smoothing spline, spectral analysis, time series.

\end{abstract}

\newpage
\spacingset{1.9} 
\section{Introduction}

Let $\{ y_t: t=1,\dots,n \}$ be a time series of length $n$, generated by a stationary  process with continuous marginal probability distribution function $F(y) := \Pr\{ y_t \le y\}$ and unique $\al$-quantile $q(\al) := F^{-1}(\al)$ for all $\al \in (0,1)$.
Let $R(\tau,\al)$ 
$(\tau = 0, \pm 1, \dots)$ be the autocovariance function (ACF) of the quantile-crossing process 
\eqn
u_t(\al) := \al - I(y_t \le q(\al)),
\label{lp}
\eqqn
where $I(\cdot)$ is the indicator function.
Assume that $R(\tau,\al)$ is absolutely summary 
over $\tau$ for fixed $\al$. Then, the following spectrum is well-defined:
\eqn
S(\om,\al) := \sum_{\tau = -\infty} ^\infty R(\tau,\al) \exp(-i \om \tau),
\label{S}
\eqqn
where  $\om \in (-\pi, \pi]$ and $i := \sqrt{-1}$.  This spectrum, which we call the quantile-crossing spectrum, is a scaled version of the quantile spectrum  introduced in Li (2008; 2012) through trigonometric quantile regression (Koenker 2005). The latter includes a scaling factor of the form $[\dot{F}(q(\al))]^{-2}$, where $\dot{F}(\cdot)$ denotes the marginal probability density function of $\{ y_t \}$. 

The quantile-crossing process in (\ref{lp}) is a variation of the level-crossing process $I(y_t \le y)$ for fixed $y \in \bbR$. An important example 
of level-crossing processes is the zero-crossing process $I(y_t \le 0)$. 
This process and its application in spectral analysis 
were the subject of the pioneering works of Hinich (1967), Brillinger (1968), and Kedem (1986). The work 
of Davis and Mikosch (2009) focused on the statistical behavior of the level-crossing 
process at extreme levels.

The level-crossing process  $I(y_t \le y)$ has mean $F(y)$ and variance $F(y) (1-F(y))$.
A reparameterzation of the level $y$  with the quantile level $\al$ through 
the $\al$-quantile $q(\al)$ yields the quantile-crossing process in (\ref{lp}) which has mean 0 
and variance $\al(1-\al)$. The quantile-crossing process and its spectral properties are investigated in several publications, including Linton and Whang (2007), Hagemann (2013), Dette et al.\ (2015), 
Kley et al. (2016), Birr et al.\ (2017), and Goto et al.\ (2022).

Besides its robustness against outliers and nonlinear distortions (Li 2008; Hagemann 2013), 
an important reason for recent interest in 
the quantile-crossing spectrum is its capability of representing nonlinear dynamics (Li 2012; Dette et al.\ 2015). 
To illustrate this point, let $F_\tau(y,y') := \Pr\{ y_{t} \le y, y_{t-\tau} \le y'\}$  and 
$\gam_\tau(y,y') := \Pr\{ (y_{t} - y) \, (y_{t-\tau}-y' )<0 \}$ 
be the lag-$\tau$  bivariate distribution function and level-crossing rate, respectively $(\tau = 0, \pm 1, \dots)$.  It is easy to show that
\eq
R(\tau,\al)  & = & \al(1-\al) - \frac{1}{2} \, \gam_\tau(q(\al) ,q(\al))  \\
&=&  F_\tau(q(\al),q(\al)) - \al^2.
\eqq
The quantity $\gam_\tau(q(\al),q(\al))$ in the first expression is an extension of  the zero-crossing rates  (Kedem 1986)
to all level-crossings in the range of $y_t$.
The quantity $F_\tau(q(\al),q(\al))$ in the second expression, which can be written as $F_\tau(F^{-1}(\al),F^{-1}(\al))$,
 is the bivariate copula function 
of $(y_t,y_{t-\tau})$ evaluated on the diagonal of the unit square $(0,1) \times (0,1)$ which is totally independent
of the marginal distribution. This explains why $S(\om,\al)$  in (\ref{S}) is sometimes called  the copula spectrum (Kley et al.\ 2016).  In the Gaussian case, $R(\tau,\al)$ is a nonlinear monotone function of the lag-$\tau$ autocovariance of $\{ y_t \}$ for each $\tau$ (Li 2020).
In general, as $\al$ varies in the interval (0,1), the link with the level-crossing rates and  bivariate distribution functions enables the quantile-crossing spectrum to provide a richer view of possibly nonlinear serial dependence than that offered by the ordinary spectrum which merely represents linear autocorrelation (or the second moments). 

 In the current literature, the quantile-crossing spectrum is largely treated as a one-dimensional function 
of $\om$ with $\al$ fixed at a selected quantile level. When interested in multiple quantiles, the spectrum is estimated  independently for each quantile. However, there are situations in which one would like to treat the quantile-crossing spectrum as a two-dimensional function on the entire domain $(\om,\al) \in (-\pi,\pi] \times (0,1)$. When the quantile-crossing spectrum varies smoothly with $\al$, the one-dimensional spectra 
at neighboring quantiles would be similar to each other. Leveraging this similarity in estimating 
the quantile-crossing spectrum across quantiles would potentially improve the accuracy in the same 
way as a  periodogram smoother across frequencies would for estimating a continuous spectral density 
function (Brockwell and Davis 1991, p.\ 350).

In this paper, we propose a new estimator based on the method of spline autoregression (SAR).
This estimator leverages the smoothness of the quantile-crossing spectrum in $\al$ (when this assumption is valid) and jointly estimate it across quantiles as a two-dimensional function using an AR model 
with functional parameters. The SAR method has been used to device an estimator for the quantile spectrum (Li 2025b) through a surrogate, called the quantile series, which is based on the quantile discrete Fourier transform (QDFT) derived from trigonometric quantile regression (Li 2025a). The quantile spectrum contains the information about the
marginal distribution which is ignored in the quantile-crossing spectrum. 
But the availability of quantile-crossing series circumvents the  quantile series and
makes it easier to use the SAR method to device an estimator of the quantile-crossing spectrum. It also enables a rigorous theoretical investigation of some statistical properties 
of the resulting SAR estimator.

The remainder of this paper is organized as follows. In section 2, we discuss three simple methods
for estimating the quantile-crossing spectrum: the level-by-level lag-window 
estimator and AR estimator, and the AR estimator with post-smoothing across quantiles.
In section 3, we introduce the new SAR estimator. In section 4, we discuss some statistical properties
of the SAR estimator. In section 5, we present the results of some numerical experiments which compare 
the performance of these estimators. In section 6, we outline the extension of the SAR method to multiple time series and biquantile-crossing spectrum. Concluding remarks are given in section 7. Finally, 
further technical details regarding the results in section 4 are given in Appendix I.
The R functions in the `qfa' package that implement the estimators are described in Appendix II.

\section{Estimation of Quantile-Crossing Spectrum} \label{sec:estimation}

When the quantile-crossing spectrum is considered only at a single quantile level or a few selected 
quantile levels (e.g.,  Hagemann 2013; Dette et al.\ 2015), it is straightforward to develop 
a spectral estimator: one simply applies the conventional techniques of spectral estimation 
to the observed quantile-crossing series 
\eqn
\hat{u}_t(\al) := \al - I(y_t \le \hat{q}(\al))  \quad  (t=1,\dots, n), 
\label{uhat}
\eqqn
where $\hat{q}(\al)$ is an estimate of the $\al$-quantile $q(\al)$.

For example,   a  lag-window (LW) estimator of $S(\cdot,\al)$ for fixed $\al$ takes the form (Brockwell and Davis 1991, p.\ 354)
\eqn
\hat{S}_{LW}(\om,\al)
:= \sum_{|\tau| \le M} w(\tau/M) \, \hat{R}(\tau,\al) \exp(-i  \om \tau) \quad \om \in (-\pi,\pi],
\label{lw}
\eqqn
where $\{ \hat{R}(\tau,\al): |\tau| < n\}$ is the sample ACF  of $\{ \hat{u}_t(\al) : t=1,\dots,n\}$, 
$w(\cdot)$ is a nonnegative function, and $M > 0$ is the bandwidth parameter. This nonparametric estimator 
has been used by Davis and Mikosch (2009), Hagemann (2013), and Dette et al.\ (2015), among others. 
The formula in (\ref{lw}) also serves as a way of constructing alternative spectra to complement the ordinary spectrum: it suffices to substitute the ACF in (\ref{lw}) with an alternative ACF-like function (e.g., Hong 2000; Jordanger and Tj{\o}stheim 2022; 2023).

A parametric alternative that can be used to estimate $S(\cdot,\al)$ for fixed $\al$ 
is the autoregressive (AR) estimator 
 \eqn
\hat{S}(\om,\al)  := \frac{\hat{\sig}^2(\al)}
{\displaystyle \bigg| 1 - \sum_{j=1}^p \hat{a}_j(\al) \exp(- i j \om) \bigg|^2} \quad \om  \in (-\pi,\pi],
\label{ar}
\eqqn
where $p$ is the order of the AR model and the AR parameters $\{ \hat{a}_j(\al): j=1,\dots,p\}$ and $\hat{\sig}^2(\al)$ are  estimated from $\{ \hat{u}_t(\al) : t=1,\dots,n\}$. While
the AR model  plays an important role in modern spectral analysis (Percival and Walden 1993; 
Stoica and Moses 1997) and have been used to estimate the quantile spectrum (Chen et al.\ 2021; 
Jim\'{e}nez-Var\'{o}n et al.\ 2024), it has not been explored  for estimating the quantile-crossing spectrum in the current literature.

Conventional techniques to produce the  AR parameters in (\ref{ar}) for fixed $\al$ 
include  the least-squares method (Priestley 1981, p.\ 346) 
and the Yule-Walker method (Brockwell and Davis 1991, p.\ 239). The former regresses $\hat{u}_t(\al)$
on $[\hat{u}_{t-1}(\al),\dots,\hat{u}_{t-p}(\al)]$ by least squares.
The latter solves the Yule-Walker equations of an AR$(p)$ process with $\{ \hat{R}(\tau,\al): |\tau| < n\}$ in place 
of the theoretical ACF. The AR model was advocated as a general approach to spectral estimation by prominent researchers such as  Akaike (1969) and Parzen (1969).  Indeed, an AR model of the form (\ref{ar}) is able to approximate any spectral density function  (Brockwell and Davis 1991, pp.\ 130--131).  The AR estimator also has some desired asymptotic properties including consistency and normality as $n \linfty$ and $p\linfty$ (Berk 1974). Unlike the bandwidth parameter $M$ in the LW estimator which is often chosen by ad hoc means,  the order $p$ in the AR estimator can be selected by simple data-driven criteria such as AIC, resulting in 
an automatic and effective procedure in practice.

In this paper, we will treat the AR estimator as an alternative to the LW estimator in estimating 
the quantile-crossing spectrum for fixed $\al$ in a set of quantile levels $\{ \al_\ell: \ell=1,\dots,L\}$.

In addition, we will consider the more general problem of estimating the quantile-crossing spectrum  as a bivariate function of $(\om,\al)$ in the entire domain $(-\pi,\pi] \times [\alu,\albar]$. This extends the problem of estimating the quantile-crossing spectrum as a univariate function of $\om$  for  fixed $\al$ considered by  Hagemann (2013), Dette et al.\  (2015), and Kley et al.\ (2016), among others. 

A simple way to obtain an estimate for all $\al \in [\alu,\albar]$ is to apply a nonparametric 
smoother to the sequence of parameters in the AR estimator produced independently 
for each $\al_\ell$ $(\ell=1,\dots,L)$, where $\al_1 := \alu < \al_2 < \dots < \al_L :=\albar$.
Smoothing the AR parameters  across the quantiles rather than interpolating them
has the  benefit of reducing the  statistical variability 
of the results and thereby producing a better estimate 
when the underlying spectrum is smooth in $\al$. This post-smoothing idea has been explored by Chen et al.\ (2021) and Jim\'{e}nez-Var\'{o}n et al.\ (2024) for estimating the quantile spectrum derived from trigonometric quantile regression. It is straightforward to adopt this idea to estimate the quantile-crossing spectrum. We will refer to the resulting estimator as the AR-S estimator.

To describe the AR-S estimator more precisely, let $\{ \tilde{a}_j(\al_\ell): j=1,\dots,p\}$ 
and $\tilde{\sig}^2(\al_\ell)$ denote the AR parameters obtained by fitting an AR$(p)$ 
model to $\{ \hat{u}_t(\al_\ell): t=1,\dots,n\}$ using the least-square or Yule-Walker method for each fixed $\al_\ell$ $(\ell = 1,\dots,L)$. The AR-S estimator is obtained by smoothing the sequences
$\{ \tilde{a}_j(\al_\ell): \ell=1,\dots,L\}$ $(j=1,\dots,p)$ 
and $\{ \tilde{\sig}^2(\al_\ell): \ell=1,\dots,L\}$ using 
 the smoothing spline method (Hastie and Tibshirani 1990, p.\ 27), which  yields
\eqn
\hat{a}_j(\cdot) := \operatorname*{argmin}_{a_j(\cdot)  \in \cF}
\bigg\{
\sum_{\ell=1}^L [ \tilde{a}_j(\al_\ell) - a_j(\al_\ell) ]^2
+ \lam_j \int_{\alu}^{\albar} [\ddot{a}_j(\al)]^2 \, d\al \bigg\}
\label{ars}
\eqqn
for $j=1,\dots,p$ and
\eqn
\hat{\sig}^2(\cdot)  :=  \operatorname*{argmin}_{\sig^2(\cdot) \in \cF} \bigg\{ \sum_{\ell=1}^L 
[ \tilde{\sig}^2(\al_\ell) - \sig^2(\al_\ell)]^2 + \  \lam_\sig \,  \int_{\alu}^{\albar} [\ddot{\sig}^2(\al)]^2 \, d\al \bigg\},
\label{sig}
\eqqn
where $\cF$ denotes the functional space of splines on $[\alu,\albar]$,
$\lam_j > 0$ and $\lam_\sig > 0$ are the smoothing parameters.   A popular method for selecting 
$\lam_j$ and $\lam_\sig$ is to use the generalized cross-validation (GCV) criterion (Hastie and Tibshirani 1990, p.\ 49).

To specify the order $p$, we first  compute the standard AIC criterion, $\text{AIC}_p(\al_\ell)$,
for each $\al_\ell$ $(\ell=1,\dots,L)$ and  $p=0,1,\dots,p_{\max}$, where $p_{\max}$ is a predetermined maximum order; 
then we choose the final $p$ in (\ref{ar}) as the minimizer of the average $L^{-1} \sum_{\ell=1}^L \text{AIC}_p(\al_\ell)$ over $ \{0,1,\dots,p_{\max}\}$.
This criterion is a comprise of potentially different requirements on the order parameter across quantiles.
Using different orders for different quantiles may introduce discontinuities in the estimated spectrum across quantiles.

In the next section, we employ the SAR method to device an alternative to the AR-S estimator.   Instead of separating function estimation from autoregression, this new  estimator combines them into a single procedure where  functional AR parameters are fitted directly to the time series by penalized least-squares autoregression.

\section{Spline Autoregression Estimator}

We assume that $S(\om,\al)$ is a suitably smooth function not only in $\om$ (as implied by the absolute 
summability of $R(\tau,\al)$ over $\tau$ for fixed $\al$) but also in $\al$. When $\{ y_t \}$ is an $m$-dependent process for some $ m \ge 1$, the smoothness of $S(\om,\al)$ in $\al$ is guaranteed by the smoothness of 
the bivariate distribution function $F_\tau(y,y)$ in $y$ and hence the smoothness of $R(\tau,\al)$ in $\al$ 
for $\tau=1,...,m$. More generally, $S(\om,\al)$ is continuous in $\al$ if (a) 
$F_\tau(y,y)$ is a continuous function of $y$ for every $\tau$ so that $R(\tau,\al)$ is a continuous  function of $\al$, 
for every $\tau$,  and (b) $|R(\tau,\al)|$ is uniformly summable over $\tau$. Furthermore, when $F_\tau(y,y)$ is $k$ times continuously differentiable in $y$ for every $\tau$ and $q(\al)$ is $k$ times differentiable in $\al$, then $S(\om,\al)$ is 
$k$ times continuously differentiable in $\al$ if the derivatives of $R(\tau,\cdot)$ up to order $k$ are  uniformly absolutely summary over $\tau$. 

By extending the least-squares method of autoregression to include functional coefficients,
the SAR method produces $\hat{a}_j(\cdot)$ $(j=1,\dots,p)$ in  (\ref{ar}) 
as functions in $[\alu,\albar]$ by
\eqn
\{ \hat{a}_1(\cdot),\dots,\hat{a}_p(\cdot)\} & := &
\operatorname*{argmin}_{a_1(\cdot), \dots, a_p(\cdot) \in \cF} \bigg\{ (n-p)^{-1} 
\sum_{\ell=1}^L\sum_{t=p+1}^n
\bigg[ \hat{u}_t(\al_\ell) - \sum_{j=1}^p a_j(\al_\ell) \, \hat{u}_{t-j}(\al_\ell) \bigg]^2  \notag \\
& & + \ \lam \, 
\sum_{j=1}^p   \int_{\alu}^{\albar}  [\ddot{a}_j(\al)]^2\, d\al \bigg\}.
\label{sar3}
\eqqn
This is a penalized least-squares problem for autoregression with spline coefficients, hence the name SAR. 
In this problem, the roughness of the functional AR coefficients is measured by 
the integral of squared second derivative and control by the smoothing parameter $\lam > 0$.
It is inspired by the smoothing spline method for nonparametric function estimation, which also has a Bayesian interpretation (Wahba 1990, p.\ 16). In the extreme case of $\lam \lzero$, the SAR solution becomes
an interpolator of the AR estimates. With $\lam > 0$ in general, the SAR solution moves away from being an interpolator and produces a smoother result that balances the goodness-of-fit with the roughness penalty.

In addition, as with the AR-S estimator discussed in the previous section, let $\hat{\sig}^2(\cdot)$ be given by (\ref{sig}) 
and let $p$ be the minimizer of the average AIC across quantiles. 
Substituting the resulting $\{ \hat{a}_j(\cdot): j=1,\dots,p\}$ and $\hat{\sig}^2(\cdot)$ in (\ref{ar}) 
gives the proposed SAR estimator of $S(\om,\al)$ for all $(\om,\al) \in (-\pi,\pi] \times [\alu,\albar]$.

The SAR problem (\ref{sar3}) has a closed-form  solution. Indeed, 
let  $\{ \phi_k(\cdot): k=1,\dots,K\}$ be a set of basis functions 
of $\cF$ so that any function $a_j(\cdot)$ in $\cF$ can be expressed as 
$a_j(\cdot)  := \sum_{k=1}^K  \theta_{jk} \, \phi_k(\cdot) = \bmphi^T(\cdot) \, \bmth_j$ for some 
$\bmth_j := [\theta_{j1},\dots,\theta_{jK}]^T \in \bbR^K$, where
\eq
\bmphi(\al)  :=  [ \phi_1(\al),\dots,\phi_K(\al)]^T \in \bbR^{K}.
\eqq
 Define 
\eq
\bmth :=  [\bmth_{1}^T,\dots,\bmth_{p}^T]^T \in \bbR^{Kp}, \quad
\bPhi(\al)  :=   \bI_p \otimes \bmphi(\al)  \in \bbR^{Kp \times p},
\eqq
where $\otimes$ stands for the Kronecker product. Then, we can write
\eq
\ba(\cdot) := [a_1(\cdot),\dots,a_p(\cdot)]^T = \bPhi^T(\cdot)\, \bmth.
\eqq
Moreover, let
\eq 
\bu_\ell & := & [\hat{u}_{p+1}(\al_\ell),\dots,\hat{u}_n(\al_\ell)]^T  \in \bbR^{n-p}, \\
\bU_\ell  & := &
\left[
\begin{array}{ccc}
  \hat{u}_p(\al_\ell) & \cdots &  \hat{u}_{1}(\al_\ell) \\
\vdots &         & \vdots \\
  \hat{u}_{n-1}(\al_\ell)  & \cdots &  \hat{u}_{n-p}(\al_\ell)
\end{array}
\right]  \in \bbR^{(n-p) \times p}, \\
\bX_\ell  & := & \bU_\ell \, \bPhi^T(\al_\ell)  \in \bbR^{(n-p) \times Kp},
\eqq
and
\eqn
\bQ   :=   \bigg[ \int_{\alu}^{\albar}
\ddot{\phi}_k(\al) \ddot{\phi}_{k'}(\al)  \, d\al\bigg]_{k,k'=1}^K \in \bbR^{K \times K}. 
\label{Phi2}
\eqqn
Then, we can write
\begin{gather}
\sum_{t=p+1}^n
\bigg[ \hat{u}_t(\al_\ell) - \sum_{j=1}^p  \, a_j(\al_\ell) \, \hat{u}_{t-j}(\al_\ell)  \bigg]^2 
 = \| \bu_\ell - \bU_\ell \, \ba(\al_\ell)  \|^2 =  \| \bu_\ell -\bX_\ell \bmth \|^2,
\label{eq1} \\
\sum_{j=1}^p \int_{\alu}^{\albar}  [\ddot{a}_j(\al)]^2\, d\al  
= \sum_{j=1}^p \bmth_j^T \bQ \, \bmth_j = \bmth^T (\bI_p \otimes \bQ) \, \bmth.
\label{eq2}
\end{gather}
Substituting (\ref{eq1}) and (\ref{eq2}) in (\ref{sar3}) leads to a reformulated SAR problem:
\eqn
\hat{\bmth}  := 
\operatorname*{argmin}_{\bmth \in \bbR^{Kp}} \ \bigg\{
\sum_{\ell=1}^L  \| \bu_\ell -\bX_\ell \bmth \|^2  +  (n-p) \lam   \bmth^T (\bI_p \otimes \bQ)  \,\bmth \bigg\}.
\label{sar3b}
\eqqn
Because the normal equations of (\ref{sar3b}) take the form
\eq
\bigg( \sum_{\ell=1}^L \bX_\ell^T \bX_\ell + (n-p) \lam   (\bI_p \otimes \bQ)  \bigg) \bmth
= \sum_{\ell=1}^L \bX_\ell^T \bu_\ell,
\eqq
 the solution $\hat{\bmth} := [\hat{\bmth}_1,\dots,\hat{\bmth}_p]^T$ in (\ref{sar3b}) can be expressed as
\eqn
\hat{\bmth} =  \bigg( \sum_{\ell=1}^L \bX_\ell^T \bX_\ell + (n-p) \lam   (\bI_p \otimes \bQ) \bigg)^{-1} 
\bigg(\sum_{\ell=1}^L \bX_\ell^T \bu_\ell \bigg).
\label{sol1}
\eqqn
Therefore, the solution to (\ref{sar3}) is given by
\eqn
\hat{\ba}(\cdot) :=  [\hat{a}_1(\cdot),\dots,\hat{a}_p(\cdot)]^T =
\bPhi^T(\cdot) \, \hat{\bmth},
\label{ahat}
\eqqn
where $\hat{a}_j(\cdot) := \bmphi^T(\cdot) \, \hat{\bmth}_j$ $(j= 1,\dots,p)$.

To address the issue of selecting the smoothing parameter $\lam$ in (\ref{sar3}), 
it is beneficial to observe that the hat matrix (or smoothing matrix) associated with the solution in (\ref{sol1}) is given by
\eq
\bH_\lam  :=  \bX_0 \bigg( \sum_{\ell=1}^L \bX_\ell^T \bX_\ell + (n-p) \lam  (\bI_p \otimes \bQ)   \bigg)^{-1} 
\bX_0^T,
\eqq
where  $\bX_0 := [\bX_1^T,\dots,\bX_L^T]^T$. Therefore,
\eqn
\tr(\bH_\lam) =  \sum_{\ell'=1}^L  \tr \bigg( \bX_{\ell'} \bigg( \sum_{\ell=1}^L \bX_\ell^T \bX_\ell 
+ (n-p) \lam (\bI_p \otimes \bQ)  \bigg)^{-1} \bX_{\ell'}^T \bigg).
\label{H}
\eqqn
The GCV criterion for selecting $\lam$ can be expressed as
\eqn
\text{GCV}(\lam) 
= 
\frac{ (L(n-p))^{-1} \sum_{\ell=1}^L  \| \bu_\ell - \bU_\ell \, \hat{\ba}_\lam(\al_\ell)\|^2}
{\{1- (L (n-p))^{-1} \tr(\bH_\lam)\}^2}.
\label{gcv}
\eqqn
Here, we insert the subscript $\lam$ in $\hat{\ba}(\cdot)$ given by (\ref{ahat}) to emphasize its dependence on $\lam$. The optimal value of $\lam$ minimizes the GCV in (\ref{gcv}) and can be found numerically 
by a line search procedure such as the R function {\tt optim} (R Core Team 2024).

It is worth noting that  an attempt of using the GCV in (\ref{gcv}) to select the order $p$ together 
with $\lam$ turns out to be unfruitful due to the interplay between these parameters in 
$\tr(\bH_\lam)$. This quantity serves as the effective degree of freedom in the SAR problem, similarly to 
its counterpart in the general problem of function estimation  (Hastie and Tibshirani 1990, p.\ 52). It is 
observed that the response of $\tr(\bH_\lam)$ to an increase in $p$ (less smoothing with 
respect to $\om$) can be mitigated by an increase in $\lam$ (more smoothing with respect to $\al$). 
For this reason, we choose to select $p$ in advance as discussed in the previous section using the AIC criterion.

\section{Characterization of the SAR Estimator}

In this section, we examine the limiting value of the SAR estimator defined by (\ref{ar}), (\ref{sig}), and
(\ref{sar3}) as the sample size grows without bound and the smoothing parameters approach zero.

Toward that end,  define
\eqn
\bGam_p(\al) :=  [R(|j-j'|,\al)]_{j,j'=1}^p, 
\quad \bmgam_p(\al)  :=  [R(1,\al),\dots,R(p,\al)]^T. \label{Gam}
\eqqn
If $\bGam_p(\al)$ is positive-definite for all $\al \in (0,1)$, then the Yule-Walker equations 
$\bGam_p(\al) \, \ba =  \bmgam_p(\al)$ have a unique solution 
\eqn
\ba_p(\al)  := [ a_{p1}(\al),\dots, a_{pp}(\al)]^T :=  \bGam_p^{-1}(\al) \, \bmgam_p(\al),
\label{ar02}
\eqqn
and the corresponding residual variance can be expressed as 
\eqn
\sig_p^2(\al) := R(0,\al) - \ba_p^T(\al) \, \bGam_p(\al)  \, \ba_p(\al).
\label{var0}
\eqqn
These parameters define an AR spectrum
\eqn
S_p(\om,\al) := \frac{\sig_p^2(\al)}
{\displaystyle \bigg| 1 - \sum_{j=1}^p a_{pj}(\al) \exp(- i j \om) \bigg|^2}.
\label{spec0}
\eqqn
For each fixed $\al \in (0,1)$, $S_p(\cdot,\al)$ maximizes the spectral entropy among all 
spectral densities for which the first $p+1$ autocovariances 
coincide with $\{ R(\tau,\al): \tau=0,1,\dots,p\}$  (Brockwell and Davis 1991, p.\ 191). For this reason, 
we refer to $S_p(\cdot,\al)$ in (\ref{spec0}) as
the maximum entropy spectrum of $\{ u_t(\al) \}$.  The spectral entropy is a measure of difficulty in linear prediction of the underlying process (Brockwell and Davis 1991, p.\ 191).
As a bivariate function,  $S_p(\om,\al)$ can be viewed as an approximation to the quantile-crossing spectrum $S(\om,\al)$ in (\ref{S}) under the maximum entropy framework.
Let 
\eq
R_p(\tau,\al) := (2\pi)^{-1} \int_{-\pi}^\pi S_p(\om,\al) \, \exp(i\tau\om) \, d\om
\eqq
be the ACF associated with $S_p(\om,\al)$. Because $R_p(\tau,\al) = R(\tau,\al)$ for $|\tau| \le p$, we have
\eq
\sup_{\om \in (-\pi,\pi]} |S_p(\om,\al) - S(\om,\al)| \le \sum_{|\tau| > p} |R_p(\tau,\al)| 
+\sum_{|\tau| > p}  |R(\tau,\al)|.
\eqq 
Therefore, the approximation error tends to zero uniformly  as $p \linfty$ 
if  $|R_p(\tau,\al)|$ and $|R(\tau,\al)|$ are uniformly summable over $\tau$.

We assume that the quantile-crossing series 
$\{ \hat{u}_t(\al): t=1,\dots,n\}$ is ergodic in second moments. 
In other words, for any $\tau = 0,1,\dots, n-1$ and $\al \in (0,1)$, let 
\eq
\hat{R}(\tau,\al) := n^{-1} \sum_{t=\tau+1}^n \hat{u}_t(\al) \, \hat{u}_{t-\tau}(\al)
\eqq
be the sample lag-$\tau$ autocovariance of $\{ \hat{u}_t(\al): t=1,\dots,n\}$.
Then, the ergodicity assumption can be stated as follows:
\begin{enumerate}
\item[(A0)] $\hat{R}(\tau,\al) \plim R(\tau,\al)$ for any fixed $\tau \in \{ 0,1,\dots\}$ 
and $\al \in [\alu,\albar]$ as $n \linfty$.
\end{enumerate}
Furthermore,  let us assume the functional parameters in (\ref{spec0}) 
are members of $\cF$. Under these conditions, $S_p(\om,\al)$ becomes the limiting value of 
$\hat{S}(\om,\al)$ as $n \linfty$, $\lam \lzero$, and $\lam_\sig \lzero$.
The complete assertion is stated  in the following  theorem.

\begin{thm}
Let $\hat{S}(\om,\al)$ be defined by (\ref{ar}), (\ref{sig}), and (\ref{sar3}). 
Assume that\/ $\bGam_p(\al)$ is positive-definite for all $\al \in [\alu,\albar]$.
Assume also that there exist $\bmth_p \in \bbR^{Kp}$ and $\bmxi_p \in \bbR^K$ such that
$\ba_p(\al) = \bPhi^T(\al) \, \bmth_p$ and $\sig_p^2(\al) = \bmphi^T(\al) \, \bmxi_p$
for all $\al \in [\alu,\albar]$ and that the denominator of $S_p(\om,\al)$ in (\ref{spec0}) is 
strictly positive  for all $(\om,\al) \in (-\pi,\pi] \times [\alu,\albar]$. If, in addition, (A0) is true, then
$\hat{S}(\om,\al) \plim S_p(\om,\al)$ uniformly in $(\om,\al) \in (-\pi,\pi] \times [\alu,\albar]$ 
as $n \linfty$, $\lam \lzero$, and $\lam_\sig \lzero$.
\label{thm:abar}
\end{thm}

\noindent
{\it Proof}.  First, observe that
\eq
\bX_\ell^T \bu_\ell  =  \bPhi(\al_\ell) \bU_\ell^T \bu_\ell, \quad 
\bX_\ell^T \bX_\ell   =   \bPhi(\al_\ell) \bU_\ell^T \bU_\ell \bPhi^T(\al_\ell).
\eqq
As $n \linfty$, the assumption (A0) implies
\eqn
\quad
(n-p)^{-1} \bU_\ell^T \bu_\ell 
\plim \bmgam_p(\al_\ell), \quad
(n-p)^{-1} \bU_\ell^T   \bU_\ell
\plim \bGam_p(\al_\ell),
\label{ggam}
\eqqn
where $\bmgam_p(\al_\ell)$ and $\bGam_p(\al_\ell)$ are defined by  (\ref{Gam}). 
Therefore,
\eq
(n-p)^{-1} \bX_\ell^T \bu_\ell 
& \plim & \bPhi(\al_\ell) \, \bmgam_p(\al_\ell), \\
(n-p)^{-1} \bX_\ell^T \bX_\ell  & \plim &   \bPhi(\al_\ell) \, \bGam_p(\al_\ell) \, \bPhi^T(\al_\ell) . 
\eqq
Combining these expressions with (\ref{sol1}) gives
\eq
\hat{\bmth} 
\plim 
\bigg( \sum_{\ell=1}^L \bPhi(\al_\ell) \, \bGam_p(\al_\ell) \, \bPhi^T(\al_\ell) \bigg)^{-1}   \bigg( \sum_{\ell=1}^L  \bPhi(\al_\ell) \,   \bmgam_p(\al_\ell) \bigg)
\eqq
as $n \linfty$ and $\lam \lzero$. Combining this result with (\ref{ahat}) and (\ref{ar02}) yields
\eqn
\hat{\ba}(\al) \plim \bPhi^T(\al) \, 
\bigg( \sum_{\ell=1}^L \bPhi(\al_\ell) \, \bGam_p(\al_\ell) \, \bPhi^T(\al_\ell) \bigg)^{-1}   \bigg( \sum_{\ell=1}^L  \bPhi(\al_\ell)  \, \bGam_p(\al_\ell) \, \ba_p(\al_\ell) \bigg)
= \ba_p(\al)
\label{limit1}
\eqqn
uniformly in $\al \in [\alu,\albar]$ as $n \linfty$ and $\lam \lzero$. The last expression is true under the assumption that $\ba_p(\al) = \bPhi^T(\al)\, \bmth_p$ for all $\al \in [\alu,\albar]$.

Moreover, the residual variance from fitting 
an AR($p$) model to $\{ \hat{u}_t(\al_\ell): t=1,\dots,n\}$ for fixed $\al_\ell$ can be written as 
\eq
\tilde{\sig}^2(\al_\ell)  & = & (n-p)^{-1} \| \bu_\ell - \bU_\ell \, \tilde{\ba}(\al_\ell) \|^2 \\
& = & (n-p)^{-1}
[\bu_\ell^T \bu_\ell - 2 \bu_\ell^T \bU_\ell \, \tilde{\ba}(\al_\ell)  
+ \tilde{\ba}^T(\al_\ell) \bU_\ell^T \bU_\ell \, \tilde{\ba}(\al_\ell) ],
\eqq
where $\tilde{\ba}(\al_\ell) := [\tilde{a}_1(\al_\ell),\dots,\tilde{a}_p(\al_\ell)]^T$ is the vector of the corresponding AR coefficients.
Under  (A0), we have $(n-p)^{-1} \bu_\ell^T \bu_\ell \plim R(0,\al_\ell)$ 
and $\tilde{\ba}(\al_\ell) \plim \ba_p(\al_\ell)$ as $n \linfty$.
Combining these results with (\ref{ggam}), (\ref{ar02}), and (\ref{var0}) yields 
$\tilde{\sig}^2(\al_\ell) \plim \sig_p^2(\al_\ell)$ as $n \linfty$. On the other hand, 
\eq
\hat{\sig}_p^2(\al)  = \bmphi^T(\al) \,  
\bigg (\sum_{\ell=1}^L \bmphi(\al_\ell) \, \bmphi^T(\al_\ell)  + \lam_\sig \bQ \bigg)^{-1}  
\bigg( \sum_{\ell=1}^L \bmphi(\al_\ell) \, \tilde{\sig}^2(\al_\ell) \bigg).
\eqq
Therefore, as $n \linfty$ and $\lam_\sig \lzero$, we obtain
\eqn
\hat{\sig}_p^2(\al)  \plim \bmphi^T(\al) \,  
\bigg (\sum_{\ell=1}^L \bmphi(\al_\ell) \, \bmphi^T(\al_\ell) \bigg)^{-1}  
\bigg( \sum_{\ell=1}^L \bmphi(\al_\ell) \, \sig_p^2(\al_\ell) \bigg) = \sig_p^2(\al)
\label{limit2}
\eqqn
uniformly in $\al \in [\alu,\albar]$. The last expression is the result of the assumption that
$\sig_p^2(\al) = \bmphi^T(\al) \, \bmxi_p$ for all  $\al \in [\alu,\albar]$.

Finally, the denominator of $S_p(\om,\al)$  in (\ref{spec0}) is bounded away from zero for $(\om,\al) \in (-\pi,\pi] \times  [\alu,\albar]$ by the positivity assumption and the extreme value theorem. Combining this result
with (\ref{limit1}), (\ref{limit2}), and (\ref{ar}) proves the assertion. \qed 

\bigskip
The ergodicity assumption (A0) would be relatively easy to justify by conventional arguments (e.g., Brockwell and Davis 1991, p.\ 220)  if $\{ \hat{u}_t(\al): t=1,\dots,n \}$ were replaced by $\{ u_t(\al): t=1,\dots,n \}$. 
The difficulty is due to the presence of $\hat{q}(\al)$  in (\ref{uhat}) which takes the place 
of $q(\al)$.  In the following, we provide a set of conditions that guarantee (A0).

Toward that end, define $\psi_\al(x) := \al - I(x\le 0)$ and
\eqn
\quad g(\bz, x) := \psi_\al(z_1-x) \psi_\al(z_2-x)
\label{g}
\eqqn
for $\bz := (z_1,z_2) \in \bbR^2$ and $x \in \bbR$.
The dependence of $g(\bz,x)$ on $\al$ is omitted for convenience.
With this notation, we can write $\hat{u}_t(\al) = \psi_\al(y_t - \hat{q}(\al))$ and
\eq
\hat{R}(\tau,\al) = n^{-1} \sum_{t=\tau+1}^n g(\bz_t,\hat{q}(\al)),
\eqq
where
\eqn
\bz_t := (y_t,y_{t-\tau}).
\label{z}
\eqqn
In addition, define
\eqn
g_1(\bz,x,\del)  := \sup_{x' \in B(x,\del)} g(\bz,x'), \quad 
g_2(\bz,x,\del) :=   \inf_{x' \in B(x,\del)} g(\bz,x'),
\label{gg}
\eqqn
where $B(x,\del) := \{ x' \in \Theta: |x'-x| \le \del\}$ with $\Theta$ being a compact subset of $\bbR$. 
In this notation, the dependence on $\tau$ and  $\al$ is again omitted  for convenience.

Now consider the following assumptions:
\begin{enumerate}
\item[(A1)]  $\hat{q}(\al) \plim q(\al)$ for fixed $\al$.
\item[(A2)]  For fixed $x \in \bbR$, the process $g(\bz_t,x)$ in (\ref{g})  obeys the weak law of large numbers, i.e.,
\eq
n^{-1} \sum_{t=1}^n [ g(\bz_t,x) - \E\{ g(\bz_t,x)\}] \plim 0
\eqq
This holds implicitly for  fixed $\tau$ and $\al$.
\item[(A3)] Both $g_1(\bz_t,x,\del)$ and    $g_2(\bz_t,x,\del)$ obey the weak law of large numbers, i.e.,
\eq
n^{-1} \sum_{t=1}^n [g_1(\bz_t,x,\del) - \E\{g_1(\bz_t,x,\del)\}] \plim 0, \\
n^{-1} \sum_{t=1}^n [g_2(\bz_t,x,\del) - \E\{g_2(\bz_t,x,\del)\}] \plim 0.
\eqq
This holds implicitly for fixed $\tau$ and $\al$.
\item[(A4)] The marginal distribution function $F(y)$ and the diagonal bivariate distribution function 
$F_\tau(y,y)$ $(\tau = 1,2,\dots)$ have continuous derivatives $\dot{F}(y)$ and $\dot{F}_\tau(y,y)$, respectively, with
$\dot{F}(q(\al)) > 0$ for all $\al \in (0,1)$.
\end{enumerate}

\begin{thm}
Assumptions (A1)--(A4) together imply (A0).
\label{thm:A0}
\end{thm}

\noindent
{\it Proof}. The assertion is proved  in Appendix I.  \qed

\bigskip
The analysis in Appendix I further shows that assumptions (A1)--(A3) 
can be satisfied by stationary processes with the strong mixing property and sufficiently fast vanishing
serial dependence. By definition (Rosenblatt 1956), a stationary process
$\{ y_t \}$ is said to have the strong mixing (or $\al$-mixing) property if and only if
\eqn
m_s := \sup\{|\Pr(A\cap B)-\Pr(A)\Pr(B)|: A\in \vsig(y_t: t \le 0), B \in
\vsig(y_t: t \ge s)\} \lzero
\label{mixing}
\eqqn
as $s \linfty$, where $\vsig(y_t: t \le 0)$ and $\vsig(y_t: t \ge s)$ 
denote the $\sig$-fields generated by $\{ y_t: t \le 0\}$
and $\{ y_t: t \ge s\}$, respectively. The quantity $m_s$, for  $s > 0$, is known as
the mixing number (or mixing coefficient) at lag $s$, which measures the strength of serial dependence.
The strong mixing condition is a type of weak dependence conditions. It is met trivially by the so-called $m$-dependent processes for which $m_s \equiv 0$ for sufficiently large $s$. It is also met by certain linear processes including ARMA processes (Withers 1981; Athreya and Pantula 1986; Mokkadem 1988).

The following theorem summarizes the results in Appendix I. The summability requirement 
of $\{ m_s \}$ is fulfilled if there exists a constant $\del > 1$ such that $m_s = O(s^{-\del})$ as $s \linfty$.

\begin{thm}
Let $\{ y_t \}$ be a stationary process with the strong mixing property (\ref{mixing}) such 
that $\sum_{s=1}^\infty m_s < \infty$. If  (A4) is also true, then, assumption (A0) is satisfied.
\label{thm:A0b}
\end{thm}

\noindent
{\it Proof}. The assertion follows from Propositions~\ref{pro:A2A3} and \ref{pro:A1} in Appendix I. \qed

\section{Numerical Experiments}

In this section, we present the results of some numerical experiments with simulated data that
compare the estimation accuracy of the proposed SAR estimator against the AR, AR-S, and LW 
estimators discussed in Section~\ref{sec:estimation}. 

\begin{figure}[p]
\centering
\includegraphics[height=2.1in,angle=-90]{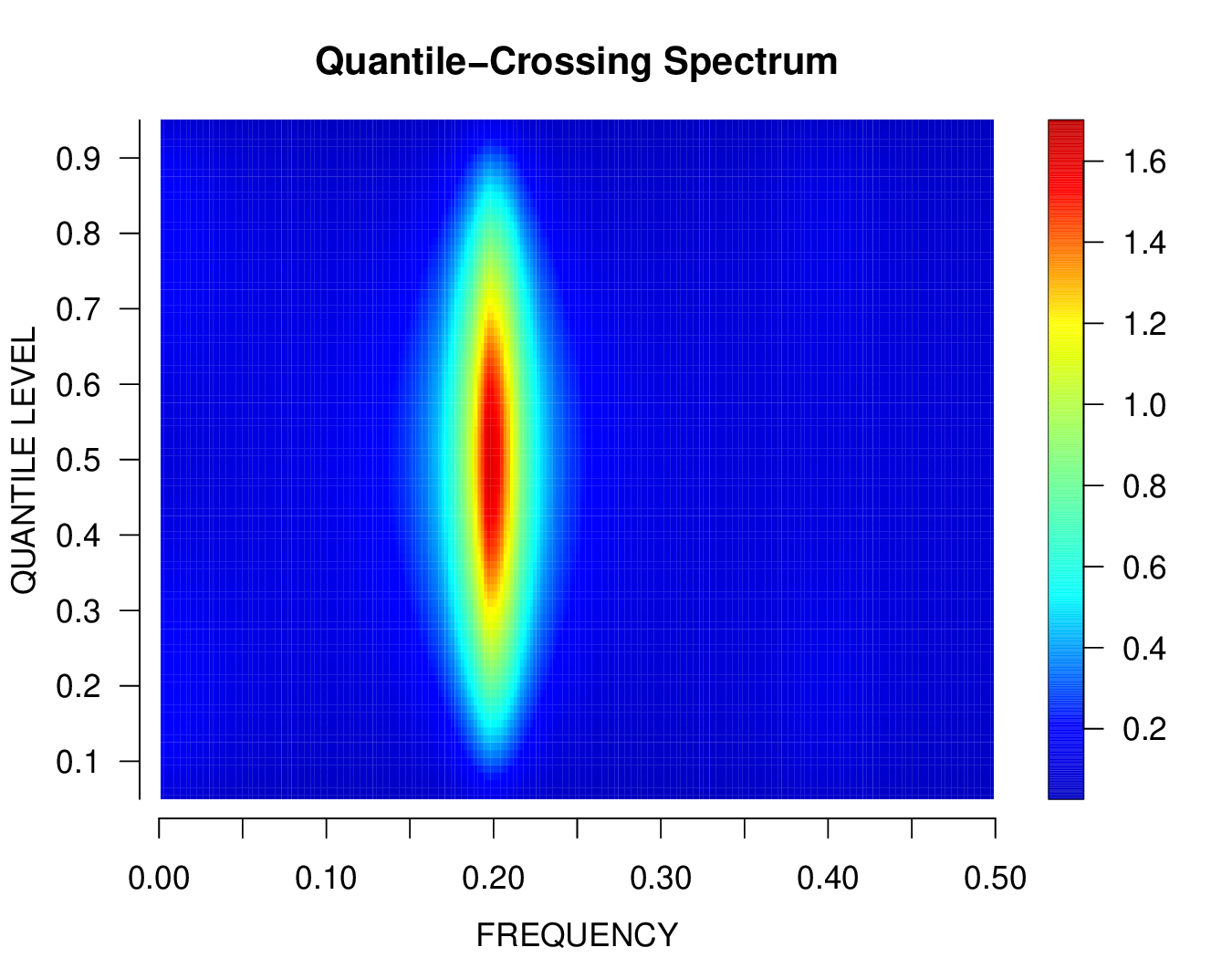}  
\includegraphics[height=2.1in,angle=-90]{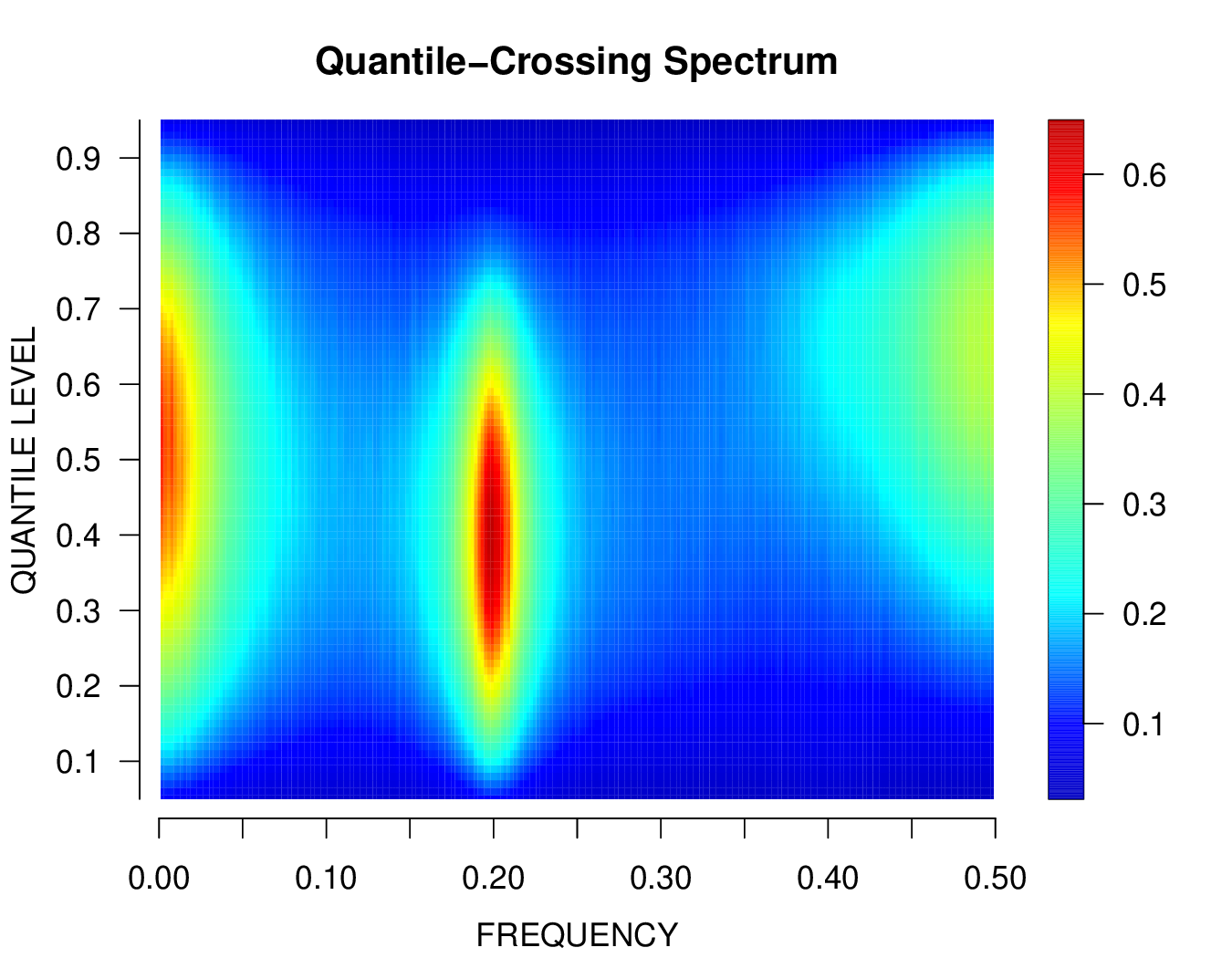} 
\includegraphics[height=2.1in,angle=-90]{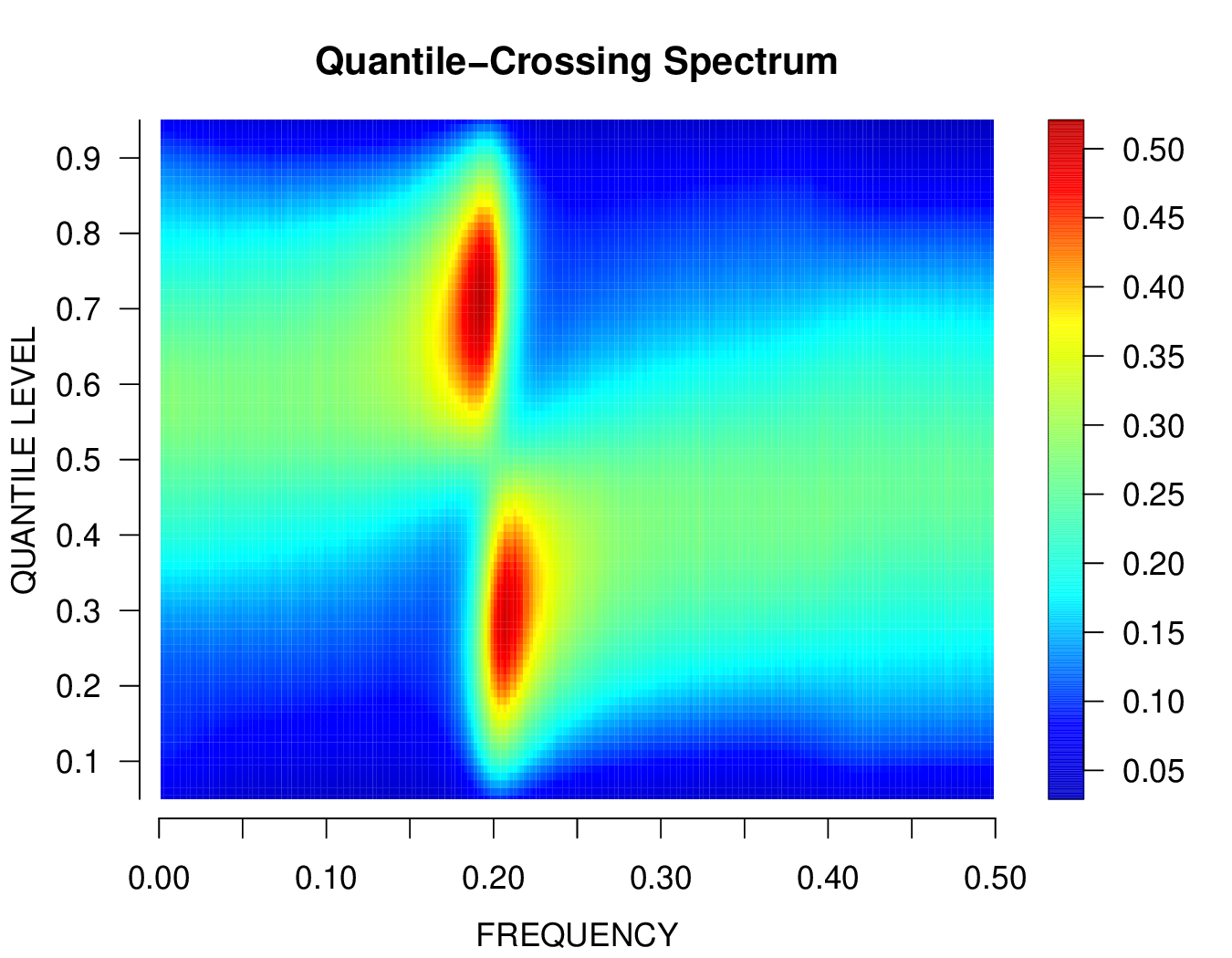} 
\center{(a) \hspace{2in}(b) \hspace{2in}(c)}   
\caption{Quantile-crossing spectra in numerical experiments: (a) case 1 with $\{ y_t \}$ given by (\ref{case1}), 
(b) case 2 with $\{ y_t \}$ given by (\ref{case2}), (c) case 3 with $\{ y_t \}$ given by (\ref{case3}). 
Horizontal axis represents the linear frequency $\om/2\pi \in (0,0.5)$.}
\label{fig:spec}
\centering
\includegraphics[height=2in,angle=-90]{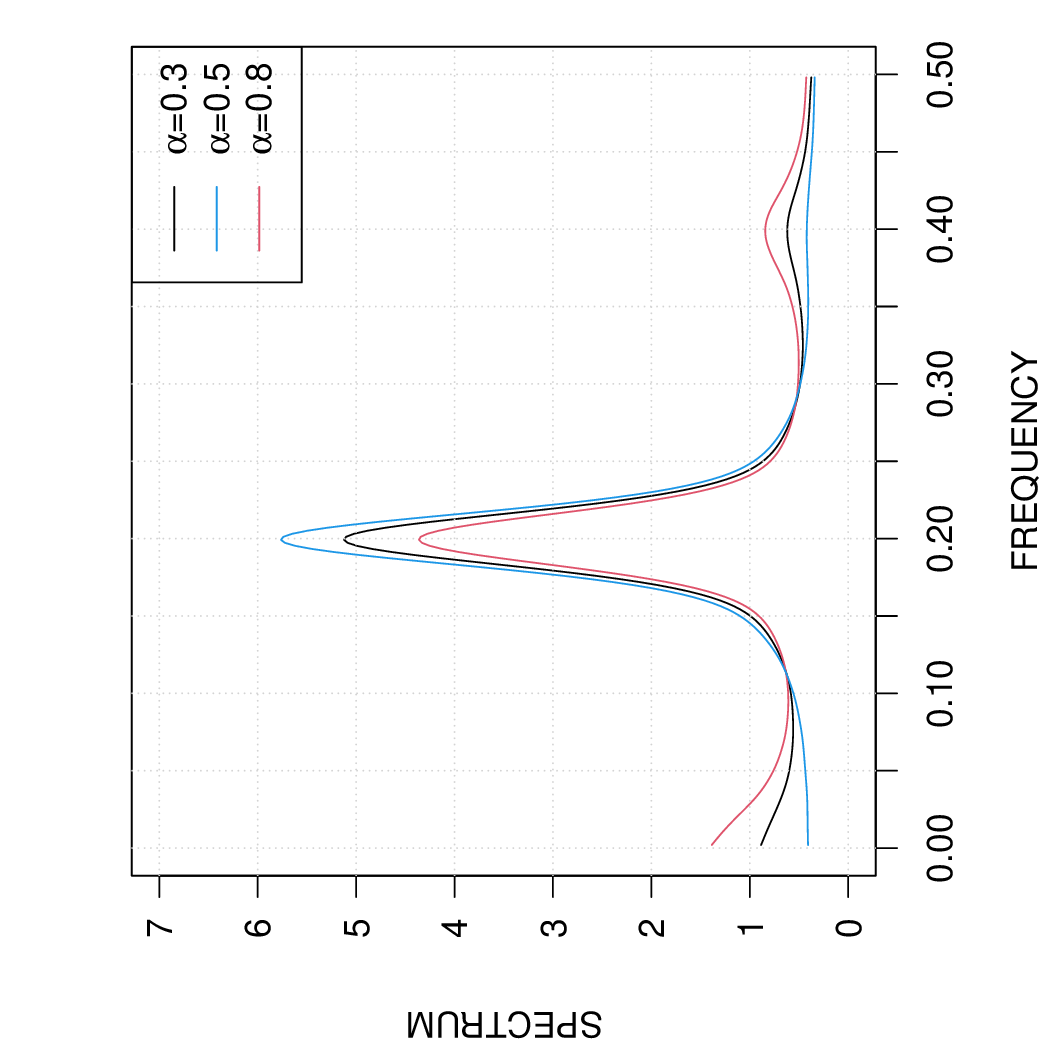}  
\includegraphics[height=2in,angle=-90]{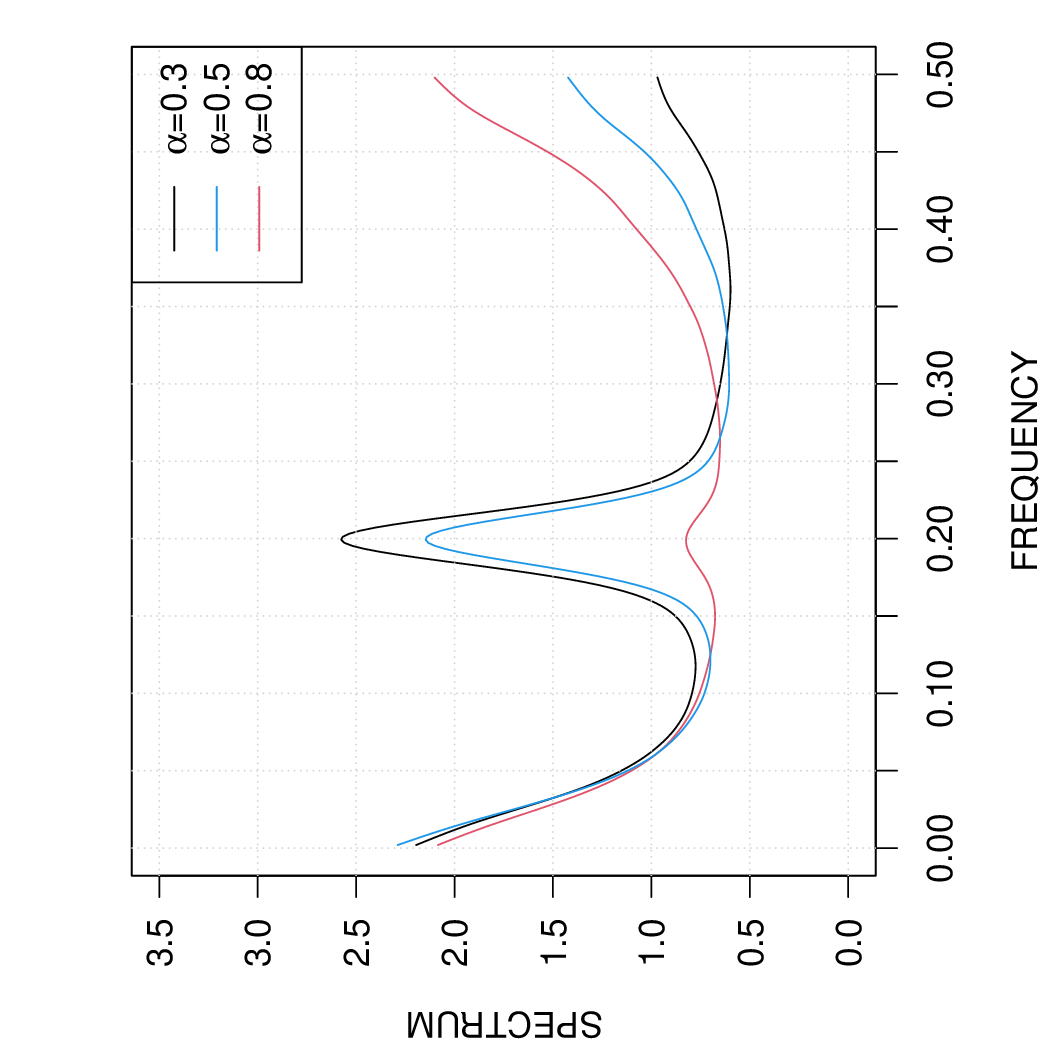} 
\includegraphics[height=2in,angle=-90]{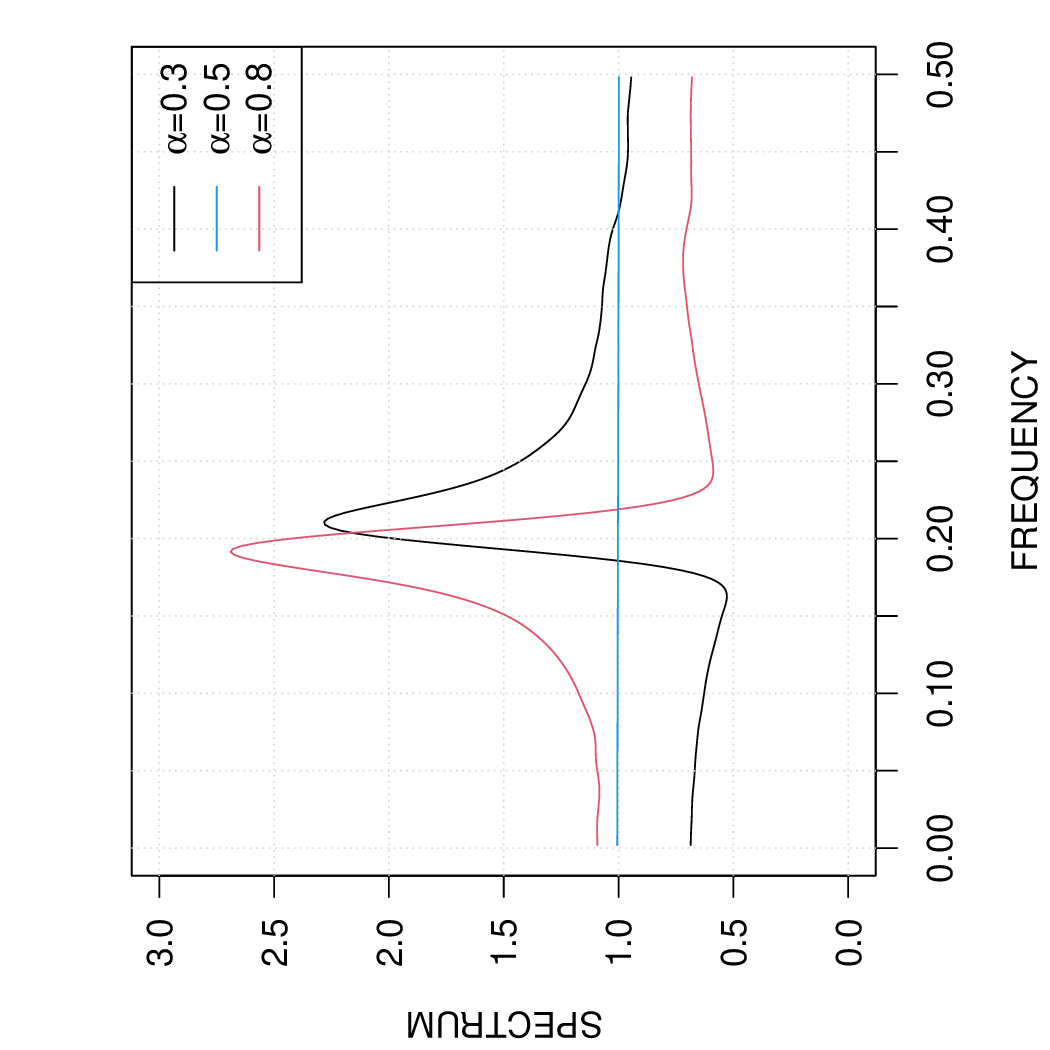} 
\center{ (a) \hspace{2in}(b) \hspace{2in}(c)}   
\caption{Quantile-crossing spectra at selected quantile levels $\al=0.3, 0.5, 0.8$: (a) case 1 with $\{ y_t \}$ given by (\ref{case1}), 
(b) case 2 with $\{ y_t \}$ given by (\ref{case2}), (c) case 3 with $\{ y_t \}$ given by (\ref{case3}). Horizontal axis represents the linear frequency $\om/2\pi \in (0,0.5)$.}
\label{fig:spec2}
\end{figure}

We compute the spectral estimates for $\om \in \{ 2\pi k /n: k=1,\dots, \lfloor (n-1)/2 \rfloor\}$ and 
$\al \in \{ \al_\ell: \ell=1,\dots,L\} = \{0.05, 0.06,\dots,0.95\}$ $(L=91)$.
Extreme quantiles are excluded because they have 
different statistical properties (Davis and Mikosch 2009). In the AR-S estimator, we smooth the 
AR parameters which are obtained independently at $\al_\ell$ $(\ell=1,\dots,L\}$ using the R function {\tt smooth.spline} 
(R Core Team 2024) with the smoothing parameters selected automatically by GCV as default. 
We also use {\tt smooth.spline}  to compute the functional residual variance in the SAR 
estimator; the smoothing parameter is set to the same value determined by the GCV criterion (\ref{gcv})
for the remaining AR parameters. In the LW estimator, we employ the Tukey-Hanning 
window (Priestley 1981, p.\ 443). We compare these estimators in three cases with the quantile-crossing spectrum having different degrees of complexity (Figures~\ref{fig:spec} and \ref{fig:spec2}).

In the first case, the time series is  an AR(2) process
\eqn
y_t = 2d \cos(2\pi f_0) \, y_{t-1} - d^2 \, y_{t-2} + \ep_t \quad (t=1,\dots,n),
\label{case1}
\eqqn
where $d = 0.9$, $f_0 = 0.2$, and $\{ \ep_t \}$ is Gaussian white noise. This process has a sharp peak around $\om /2\pi = f_0$ in its ordinary spectrum. A similar peak appears in the quantile-crossing spectrum across quantiles, as shown 
in Figures~\ref{fig:spec}(a) and \ref{fig:spec2}(a).  
The spectral peak decreases in magnitude as the quantile level moves away from 0.5 
towards  0 or 1 due to the decreasing  variance of the quantile-crossing process (\ref{lp}).

In the second case, the time series is a nonlinear mixture of three zero-mean unit-variance 
AR processes $\{ \xi_{1,t} \}$, $\{ \xi_{2,t} \}$, and $\{ \xi_{3,t} \}$, which  satisfy $\xi_{1,t} =  a_{11} \, \xi_{1,t-1} + \ep_{1,t}$,
$\xi_{2,t} =  a_{21} \, \xi_{2,t-1} + \ep_{2,t}$, and $\xi_{3,t} =  a_{31} \, \xi_{3,t-1} 
+ a_{32} \, \xi_{3,t-2} + \ep_{3,t}$,
where $a_{11}  := 0.8$, $a_{21} := -0.7$, $a_{31} := 2 d \cos(2\pi f_0)$, and 
$a_{32} := -d^2$, with $d=0.9$ and $f_0 =0.2$, and where $\{\ep_{1,t} \}$, $\{\ep_{2,t} \}$, 
and  $\{\ep_{3,t} \}$  are mutually independent Gaussian white noise. In their ordinary spectra, 
the first process $\{ \xi_{1,t} \}$ has a broad peak around $\om =0$, the second process 
$\{ \xi_{2,t} \}$ has a  broad peak around $\om = \pi$, and the third process $\{ \xi_{3,t} \}$ is identical to the process in the first case (\ref{case1}), 
with a sharp spectral peak around $\om/2\pi = 0.2$. The nonlinear mixture takes the form
\eqn
\left\{
\begin{array}{l} 
\zeta_{t}   :=  w_1(\xi_{1,t}) \, \xi_{1,t}+ (1-w_1(\xi_{1,t})) \,\xi_{2,t}, \\
y_{t}  :=  w_2(\zeta_{t}) \, \zeta_{t} + (1-w_2(\zeta_{t})) \, \xi_{3,t}, 
\end{array} \right.
\label{case2}
\eqqn
where $w_1(x) := 0.9 I(x < -0.8) + 0.2  I(x > 0.8) + \{ 0.9 - (7/16) (x + 0.8)\} I(|x| \le 0.8)$
and $w_2(x) := 0.5 I(x < -0.4) +  I(x > 0.4) + \{ 0.5 + (5/8) (x + 0.4)\} I(|x| \le 0.4)$.
The mixing function $w_1(x)$ equals 0.9 for $x < -0.8$ and 0.2 for $x > 0.8$; it interpolates these values linearly 
for $x \in [-0.8,0.8]$. Therefore, the intermediate process $\{ \zeta_t \}$ behaves more like $\{ \xi_{1,t} \}$ 
at lower quantiles and somewhat like $\{ \xi_{2,t} \}$ at higher quantiles. 
Similarly, $w_2(x)$ equals 0.5 for $x < -0.4$ and 1 for $x > 0.4$, and interpolates 
these values linearly for $x \in [-0.4,0.4]$.  Therefore, the final series $\{ y_t \}$ blends 
$\{ \zeta_t\}$ with $\{ \xi_{3,t} \}$ at lower quantiles but favors $\{ \zeta_t\}$ at higher quantiles.
 As shown in Figures~\ref{fig:spec}(b) and \ref{fig:spec2}(b), the quantile-crossing spectrum of this process 
exhibits a narrow peak around frequency 0.2, a broader peak at frequency 0, and a broader-still peak 
at frequency 0.5, which represent the respective properties of $\{ \xi_{3,t} \}$, $\{ \xi_{1,t} \}$, and $\{ \xi_{2,t} \}$. 
Unlike case 1, the peaks in case 2 are not necessarily centered symmetrically around $\al = 0.5$. 
In fact, the peak around frequency 0.2 is shifted downward, 
whereas the peak at frequency 0.5 is shifted upward.
Such a behavior is the result of judiciously designed mixing functions $w_1(x)$ and $w_2(x)$.

In the third case, we consider the stochastic volatility model studied by Hagemann (2013):
\eqn
y_t = \ep_{3,t} \exp( \xi_{3,t-1}),
\label{case3}
\eqqn
where $\{ \xi_{3,t}\}$ is the narrow-band process in case  2 and $\{ \ep_{3,t}\}$ is the associated Gaussian 
white noise. Because $\{ y_t \}$ in (\ref{case3})  is an uncorrelated series (i.e., its ACF equals zero for all nonzero lags), 
its ordinary spectrum  is flat and unable to detect the serial dependence introduced through the volatility process 
$\{ \exp(\xi_{3,t-1}) \}$. The quantile-crossing spectrum, shown in Figures~\ref{fig:spec}(c) and \ref{fig:spec2}(c), 
is able to reveal nonflat spectral patterns for a range of quantile levels. For this reason, the stochastic volatility 
model (\ref{case3}) was used in Hagemann (2013) to demonstrate how the quantile-crossing spectrum, 
estimated by the LW estimator in (\ref{lw}), can be used in a white noise test 
to detect nonlinear serial dependence.

To measure the goodness of fit for the spectral estimators, we employ the Kullback-Leibler spectral
divergence (Li 2014, p.\ 179)
\eqn
{\rm KLD} := \hat{E} \bigg\{ 
\frac{1}{L \lfloor (n-1)/2 \rfloor}  \sum_{\ell=1}^L \sum_{k=1}^{\lfloor (n-1)/2 \rfloor} 
\bigg( \frac{\hat{S}(\om_k,\al_\ell)}{S(\om_k,\al_\ell)}  - \log \frac{\hat{S}(\om_k,\al_\ell)}{S(\om_k,\al_\ell)}
-  1 \bigg) \bigg\}
\label{kld}
\eqqn
and the root mean-square error
\eqn
{\rm RMSE} := \sqrt{  \hat{E} \bigg\{  \frac{1}{L \lfloor (n-1)/2 \rfloor}  \sum_{\ell=1}^L \sum_{k=1}^{\lfloor (n-1)/2 \rfloor} 
(\hat{S}(\om_k,\al_\ell) - S(\om_k,\al_\ell) )^2\bigg\}  },
\label{rmse}
\eqqn
where $\hat{E}\{ \cdot \}$ stands for the ensemble average over independent Monte Carlo runs and 
 $\om_k := 2\pi k/n$ $(k=1,\dots, \lfloor (n-1)/2 \rfloor)$.

\begin{table}[p]
\begin{center} 
\caption{KLD and RMSE of Quantile-Crossing Spectrum Estimators}
\label{tab:err}
{\small
\begin{tabular}{l|ccc|ccc} \toprule
 \multicolumn{7}{c}{KLD} \\ \midrule
& \multicolumn{3}{c|}{$n=256$} & \multicolumn{3}{c}{$n=512$} \\
 \cmidrule{2-4} \cmidrule{5-7} 
 \multicolumn{1}{c|}{Case} &  AR    &  AR-S & SAR   & AR   & AR-S & SAR  \\  \midrule
Case 1 (\ref{case1})     &  0.0251   & 0.0242 & 0.0205 & 0.0142  & 0.0137 & 0.0118   \\
Case 2 (\ref{case2})     &  0.0318   & 0.0311 & 0.0276 & 0.0179  & 0.0174 &  0.0153 \\
Case 3 (\ref{case3})     &  0.0334   & 0.0328 & 0.0304 & 0.0212 & 0.0208 & 0.0189  \\ 
\bottomrule
\multicolumn{7}{c}{RMSE} \\ \midrule
& \multicolumn{3}{c|}{$n=256$} & \multicolumn{3}{c}{$n=512$} \\
\cmidrule{2-4} \cmidrule{5-7}
 \multicolumn{1}{c|}{Case} &  AR    & AR-S &  SAR   & AR  & AR-S  & SAR \\  \midrule
Case 1 (\ref{case1}) & 0.0668  & 0.0663 & 0.0642 &  0.0518  & 0.0515 & 0.0495 \\ 
Case 2 (\ref{case2}) & 0.0571  & 0.0567 & 0.0547 &  0.0433  & 0.0429 & 0.0412  \\
Case 3 (\ref{case3}) & 0.0525 & 0.0521 & 0.0507  &  0.0425  & 0.0422 & 0.0410 \\
\bottomrule
\end{tabular} }
\end{center}
\begin{center}
\begin{minipage}{4.4in}
{\scriptsize 
Results are based on 1000 Monte Carlo runs. }
\end{minipage}
\end{center}
\end{table}

Table~\ref{tab:err} contains these measures of  the AR, AR-S, and SAR estimators for estimating the quantile-crossing
spectrum in cases 1--3. With the order $p$ of the AR models in these estimators determined automatically by AIC and the smoothing parameter $\lam$ in SAR by GCV, the result in Table~\ref{tab:err} clearly favors the SAR estimator over the AR and AR-S estimator in terms of either KLD or RMSE.  The AR-S estimator has slightly smaller errors than the AR estimator, thanks
to the post smoothing step. However, due to the separation of smoothing from AR model fitting, 
the AR-S estimator is not as effective as the SAR estimator in these cases. In addition, 
Table~\ref{tab:err} also demonstrates that both KLD and RMSE decrease with the increase of $n$ from 256 to 512.

\begin{figure}[t]
\centering
\includegraphics[height=2in,angle=-90]{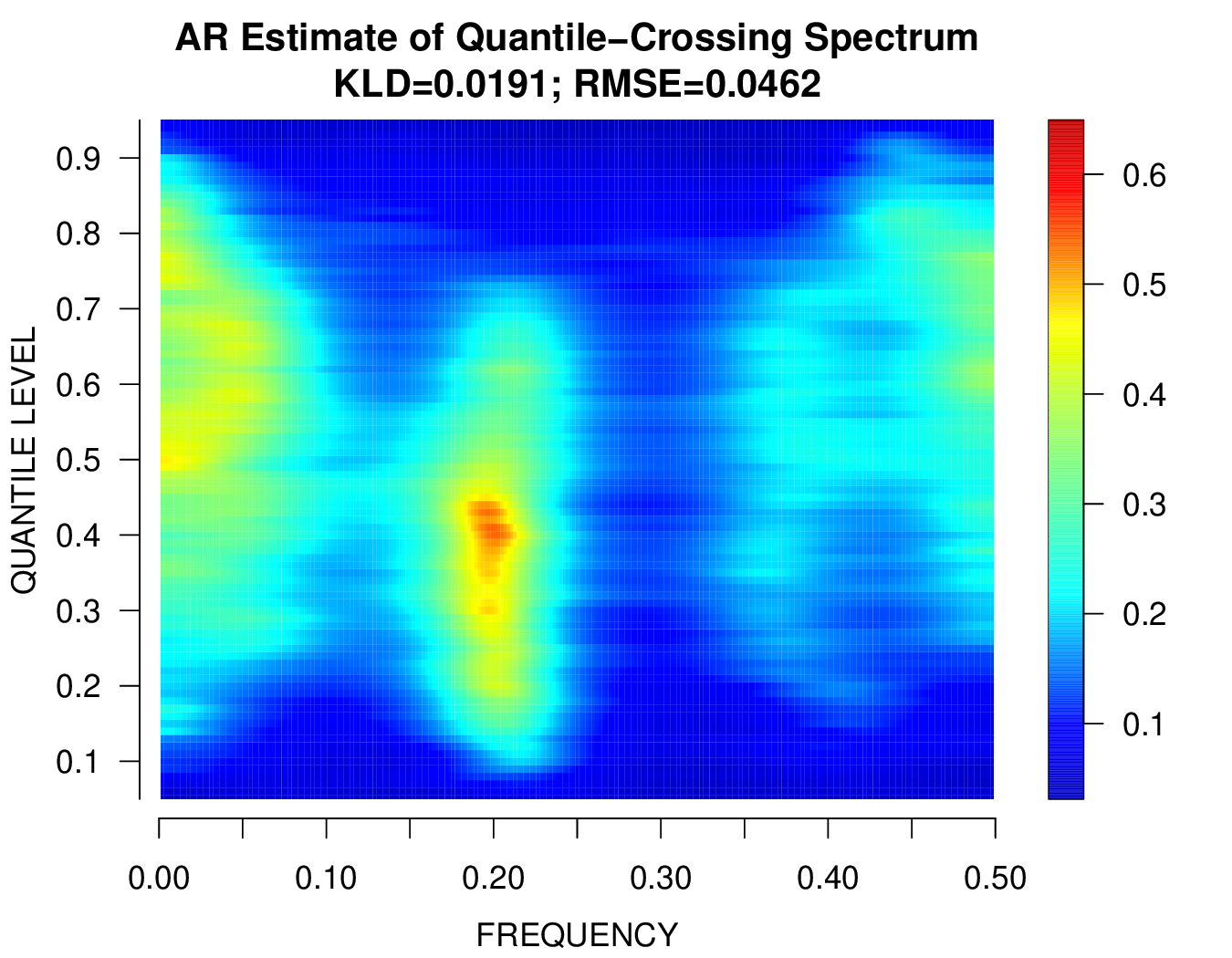}   
\includegraphics[height=2in,angle=-90]{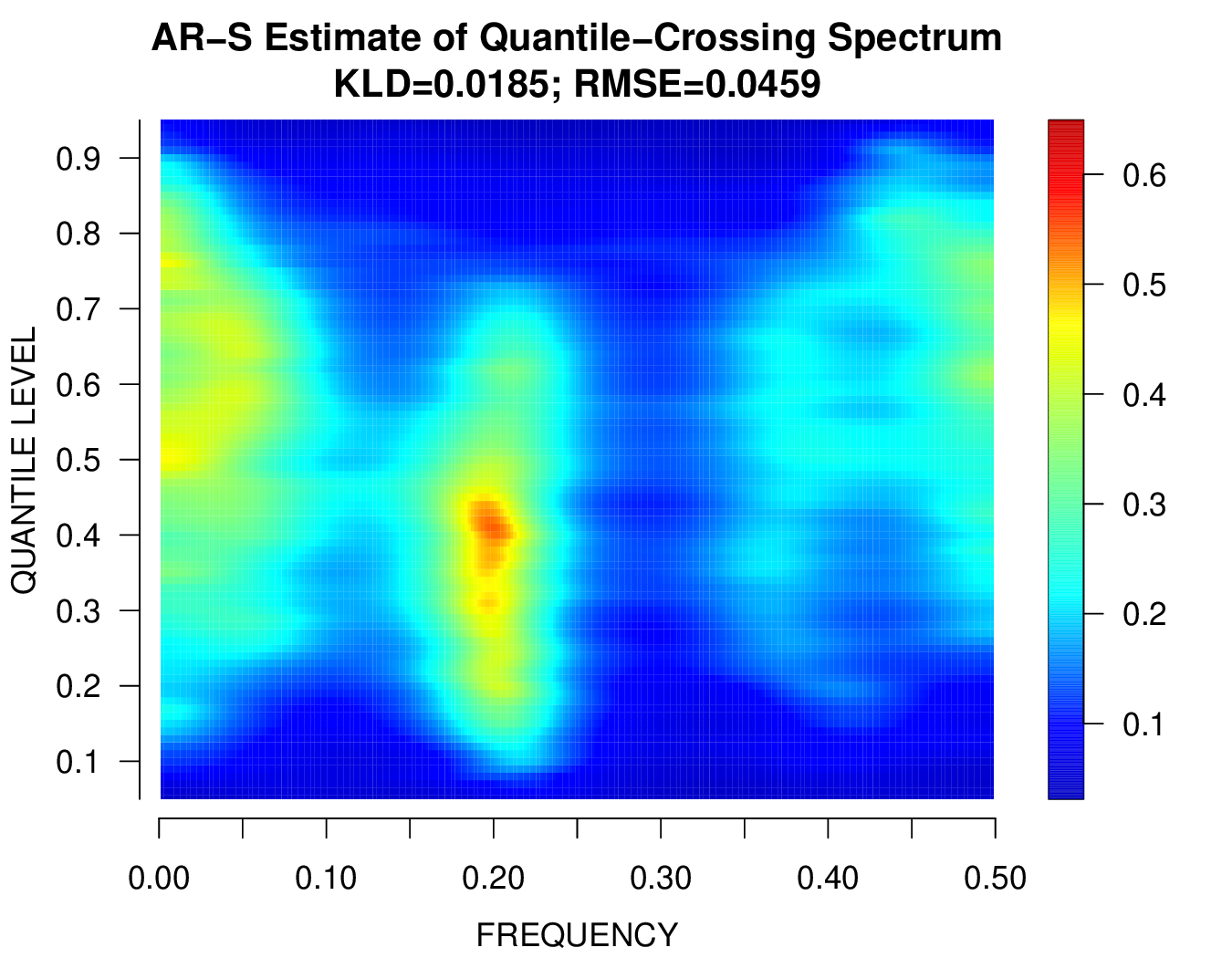}   
\includegraphics[height=2in,angle=-90]{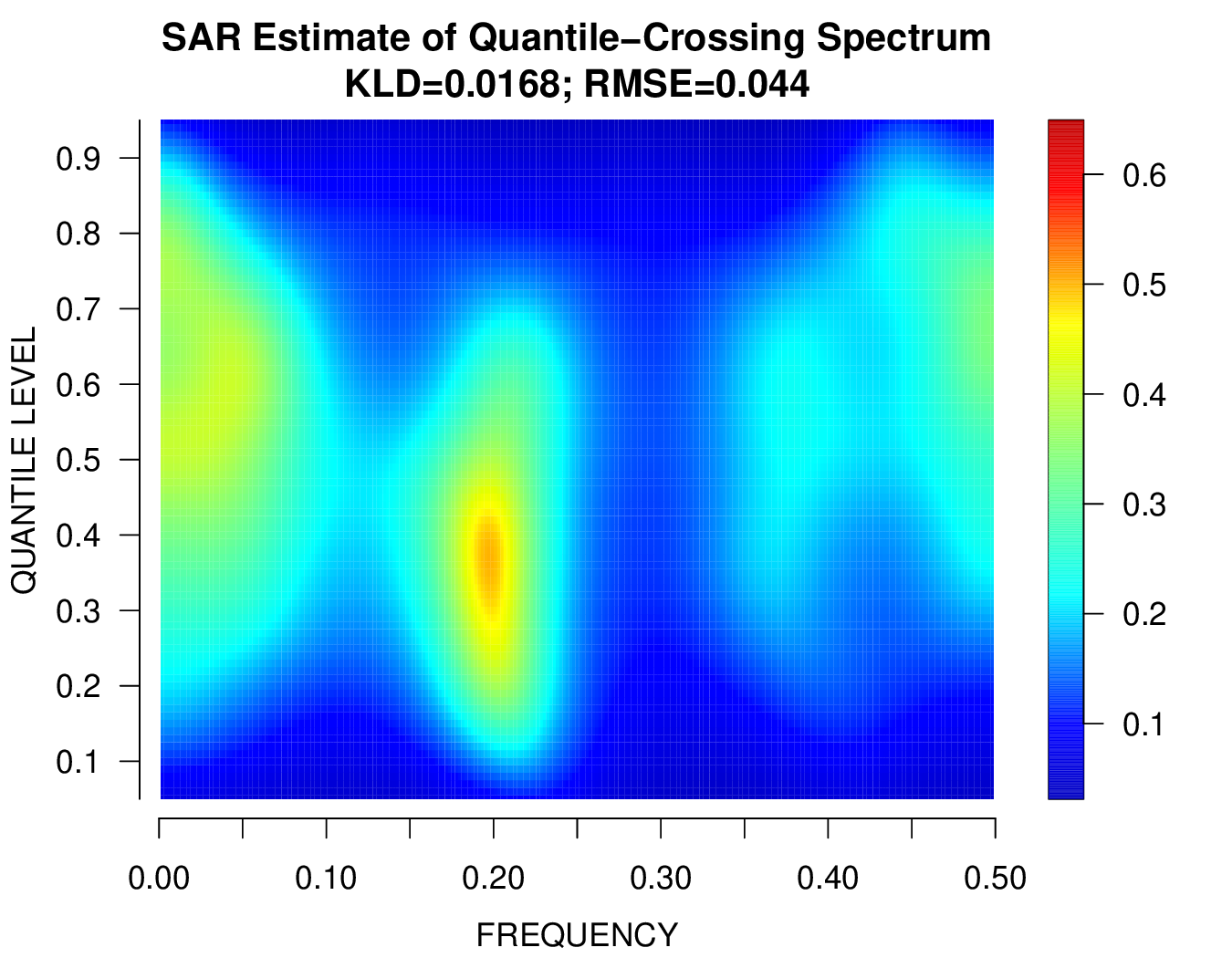}
\center{(a) \hspace{2in}(b) \hspace{2in}(c)}   
\caption{(a) AR estimate from a time series in case 2 ($n=512$): KLD = 0.0191, RMSE = 0.0462. 
(b) AR-S estimate from the same data: KLD = 0.0185, RMSE = 0.0459.
(c) SAR estimate from the same data: KLD = 0.0168, RMSE = 0.0440.}
\label{fig:sar}
\end{figure}

Figure~\ref{fig:sar} further demonstrates the effect of smoothing in SAR with respect to the quantile 
level using the estimates obtained from a time series  in case 2. Without smoothing across quantiles,
the AR estimate in Figure~\ref{fig:sar}(a) appears noisy across the quantile levels. The noise is reduced in 
the AR-S estimate (Figure~\ref{fig:sar}(c)) due to post-smoothing. The SAR estimate in Figure~\ref{fig:sar}(b) appears 
much less noisy across quantiles and more akin to the true spectrum shown in Figure~\ref{fig:spec}(b). 
The reduction of noise across quantiles is also reflected in the reduction of KLD and RMSE shown in the caption of  Figure~\ref{fig:sar}.

Finally, we compare the performance of the parametric AR estimator and the semi-parametric AR-S and SAR estimators 
with the nonparametric LW estimator.  Generally speaking, for estimating the ordinary spectrum, the LW method is known
for its flexibility in handling a variety of spectral patterns, whereas the AR method is more 
effective in capturing sharp spectral peaks. Similar behaviors are expected for these estimators in estimating
the quantile-crossing spectrum.

\begin{figure}[t]
\centering
\includegraphics[height=3in,angle=-90]{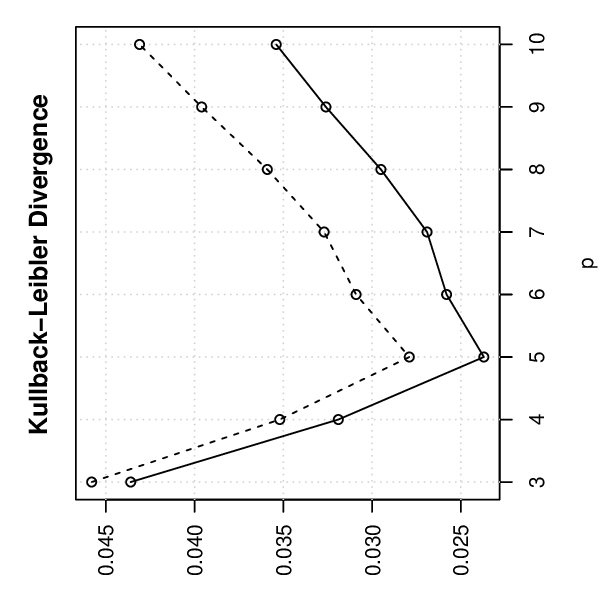}   
\includegraphics[height=3in,angle=-90]{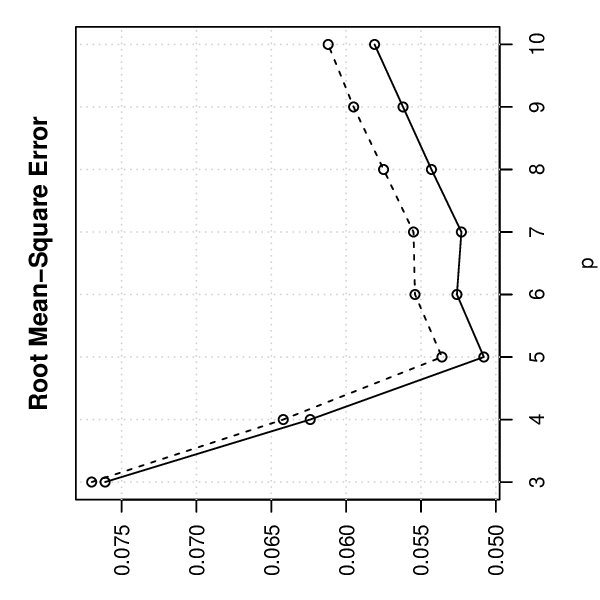}
\center{(a) \hspace{3in}(b)}   
\caption{(a) KLD and (b) RMSE of AR an SAR  estimators in case 2 ($n=256$) as a function of $p$: dashed line, AR;  
solid line, SAR. }
\label{fig:kld}
\end{figure}

Due to a lack of convenient data-driven method on par with AIC to select 
the bandwidth parameter $M$ for the LW estimator, we compare the performance in the optimal setting:  
Specifically, the bandwidth parameter $M$ in LW is fixed across Monte Carlo runs at the respective 
minimizer of the ensemble measure KLD or RMSE. Similarly, 
the order parameter $p$ in AR, AR-S, and SAR is fixed across Monte Carlo runs 
at the respective minimizer of the ensemble measure KLD or RMSE, except that the smoothing parameters 
in AR-S and SAR are still determined in each Monte Carlo run by data-driven GCV criteria. As an example, 
Figure~\ref{fig:kld} shows the ensemble measures KLD and RMSE of the AR and SAR estimators as a function of $p$ 
in case 2, where  $p=5$ is the minimizer of both KLD and RMSE for both AR and SAR.

\begin{table}[p]
\begin{center} 
\caption{KLD and RMSE of Quantile-Crossing Spectrum Estimators in Optimal Setting}
\label{tab:err2}
{\small
\begin{tabular}{l|cccc|cccc} \toprule
 \multicolumn{9}{c}{KLD}  \\ \midrule
& \multicolumn{4}{c|}{$n=256$} & \multicolumn{4}{c}{$n=512$} \\
 \cmidrule{2-5} \cmidrule{6-9}  
 \multicolumn{1}{c|}{Case} &  AR & AR-S   &  SAR   & LW & AR  & AR-S & SAR & LW \\  \midrule
Case 1 (\ref{case1})  & 0.0241 & 0.0232 & 0.0195  & 0.0350 & 0.0134 & 0.0129 & 0.0110 &  0.0218 \\
Case 2 (\ref{case2})  & 0.0279 & 0.0271 & 0.0237  & 0.0264 & 0.0165 & 0.0161 & 0.0144 &  0.0160 \\
Case 3 (\ref{case3})  & 0.0296 & 0.0289 & 0.0260  & 0.0257 & 0.0197 & 0.0192 & 0.0169 & 0.0172  \\ 
\bottomrule
 \multicolumn{9}{c}{RMSE} \\ \midrule
& \multicolumn{4}{c|}{$n=256$} & \multicolumn{4}{c}{$n=512$} \\
 \cmidrule{2-5} \cmidrule{6-9}  
 \multicolumn{1}{c|}{Case} &  AR  & AR-S &  SAR   & LW & AR  & AR-S & SAR & LW \\  \midrule
Case 1 (\ref{case1}) & 0.0654 & 0.0649 & 0.0626 & 0.0729 & 0.0500 & 0.0496 & 0.0474 & 0.0592 \\
Case 2 (\ref{case2}) & 0.0536 & 0.0531 & 0.0508 & 0.0535 & 0.0426 & 0.0408 & 0.0407 & 0.0416 \\
Case 3 (\ref{case3}) & 0.0496 & 0.0491 & 0.0474 & 0.0461 & 0.0409 & 0.0405 &  0.0389 & 0.0378 \\
\bottomrule
\end{tabular} 
}
\end{center}
\begin{center}
\begin{minipage}{5.5in}
{\scriptsize 
Results are based on 1000 Monte Carlo runs.  Ensemble minimizers are used for $p$ in AR, AR-S, and SAR and for $M$ in LW.}
\end{minipage}
\end{center}
\end{table}

Table~\ref{tab:err2} contains the resulting performance measures. 
The result shows that SAR always outperforms AR-S which, in turn, outperforms AR. 
Moreover, AR already has smaller errors than LW, which is not too surprising because the time series
is a Gaussian AR process. In cases 2 and 3, LW performs better than AR.
With the smoothing across quantiles, AR-S and SAR are able to improve the performance of AR and outperform LW in case 2, 
but the improvement is not enough to overtake LW in case 3 (the only exception is SAR for $n=512$, measured by KLD). 
Unlike case 1,  the asymmetric drop from the spectral peak 
in case 3 (Figure~\ref{fig:spec2}(c)) favors the nonparametric LW method over the parametric AR method.

\section{Extensions}

It is not difficult to extend the SAR method to multiple time series. For a vector-valued  stationary time series $\by_t := [y_{1,t},\dots,y_{m,t}]^T$, 
the quantile-crossing spectral matrix at  $\al \in (0,1)$, denoted by $\bS(\om,\al)$, is nothing but the ordinary spectrum matrix of $\bu_t(\al) :=  [u_{1,t}(\al),\dots,u_{m,t}(\al)]^T$,
where $u_{k,t}(\al)  := \al - I(y_{k,t} \le q_k(\al))$, with $q_k(\al)$ being 
the $\al$-quantile of $y_{k,t}$ $(k=1,\dots,m)$. Let $\bR(\tau,\al)$ denote 
the ACF of $\{ \bu_t(\al) \}$. If all elements in $\bR(\tau,\al)$
 are absolutely summable over $\tau$ for fixed $\al$, then
\eq
\bS(\om,\al) = \sum_{\tau=-\infty}^\infty \bR(\tau,\al) \exp(- i\om  \tau).
\eqq
The AR model of order $p$ for this spectrum takes the form
\eq
\bS_p(\om,\al) := [ \bI - \bA(\om,\al)]^{-1} \bV(\al) \, [ \bI - \bA(\om,\al)]^{-H}, 
\eqq
where $\bA(\om,\al) := \sum_{j=1}^p \bA_j(\al) \exp(-i j\om)$, with superscript $-H$ denoting
the Hermitian transpose of an inverse matrix.
When the AR parameters $\bA_1(\al),\dots,\bA_p(\al)$ and $\bV(\al)$ are obtained by solving the Yule-Walker equations formed by $\{  \bR(\tau,\al): \tau=0,1,\dots,p\}$, the resulting AR spectrum has the maximum entropy property (Parzen 1982; Choi 1993).

Let $\hat{u}_{k,t}(\al) := \al - I(y_{k,t} \le \hat{q}_k(\al))$ $(t=1,\dots,n)$, where 
$\hat{q}_k(\al)$ is the sample $\al$-quantile of $\{ y_{k,t}: t=1,\dots,n\}$. Then, the SAR problem
can be stated as follows:
\eq
\{ \hat{\bA}_1(\cdot),\dots,\hat{\bA}_p(\cdot)\}
& := & \operatorname*{argmin}_{\bA_1(\cdot), \dots, \bA_p(\cdot) \in \cF_m} 
\bigg\{ (n-p)^{-1} \sum_{\ell=1}^L\sum_{t=p+1}^n
\bigg\| \hat{\bu}_t(\al_\ell) - \sum_{j=1}^p \bA_j(\al_\ell) \, \hat{\bu}_{t-j}(\al_\ell) \bigg\|^2  \notag \\
& & + \ \lam \, 
\sum_{j=1}^p   \int_{\alu}^{\albar}  \| \ddot{\bA}_j(\al)\|^2\, d\al \bigg\},
\eqq
where $\cF_m$ denotes the space of $m$-by-$m$ matrix-valued spline functions in $[\alu,\albar]$.
The matrix $\hat{\bV}(\cdot)$ is defined as  the result of spline smoothing across the quantile levels 
applied to the residual covariance matrices of the ordinary least-squares AR fit to $\{ \hat{\bu}_t(\al_\ell):t=1,\dots,n\}$ $(\ell=1,\dots,L)$. The resulting spectral estimator takes the form
\eq
\hat{\bS}(\om,\al) := [ \bI - \hat{\bA}(\om,\al)]^{-1} \hat{\bV}(\al) \, [ \bI - \hat{\bA}(\om,\al)]^{-H},
\eqq
where $\hat{\bA}(\om,\al) := \sum_{j=1}^p \hat{\bA}_j(\al) \exp(-ij\om)$. The GCV criterion for selecting the smoothing parameter $\lam$ can be derived accordingly.

Furthermore, let $\bR(\tau,\al,\al')$ denote the cross-autocovariance function of
$\{ \bu_t(\al) \}$ and $\{ \bu_t(\al') \}$. If all elements in $\bR(\tau,\al,\al')$
 are absolutely summable over $\tau$ for fixed $\al$ and $\al'$, 
 then, we can define what may be called the biquantile-crossing spectral matrix or spectrum as
 \eq
\bS(\om,\al,\al') := \sum_{\tau=-\infty}^\infty \bR(\tau,\al,\al') \exp(- i\om  \tau).
\eqq
This spectrum has been studied in the literature  
(Dette et al.\ 2015; Kley et al.\ 2016) for fixed $\al$ and $\al'$.  It is related to the
biquantile spectrum in Li (2014, p.\ 564). It is conceivable that an SAR-based estimator of the form
\eq
\hat{\bS}(\om,\al,\al') := [ \bI - \hat{\bA}(\om,\al)]^{-1} \hat{\bV}(\al)^{1/2} \,  \hat{\bV}(\al')^{1/2}  
\, [ \bI - \hat{\bA}(\om,\al')]^{-H}
\eqq
can be employed to estimate the biquantile-crossing spectrum as a function of $\om$ in $(-\pi,\pi]$ 
for fixed $\al$ and $\al'$, or as a trivariate function
of ($\om,\al,\al')$ in $(-\pi,\pi] \times [\alu,\albar] \times [\alu',\albar']$.

\section{Concluding Remarks}

We have proposed a new estimator for estimating the quantile-crossing spectrum as a bivariate function of frequency and quantile level. This estimator is derived by the spline autoregression (SAR) method.
It jointly fits an AR model with functional coefficients to the quantile-crossing series across multiple quantiles by penalized least squares. The AR coefficients are represented as spline functions of the quantile level and penalized for their roughness. The usefulness of the proposed estimator is demonstrated by simulated examples in comparison with some parametric and nonparametric  alternatives.

The variance of the quantile-crossing process $\{ u_t(\al) \}$ equals $\al(1-\al)$ and carries
no information about the original process $\{ y_t \}$. Therefore, one may consider the normalized 
quantile-crossing process $u_t(\al)/\sqrt{\al(1-\al)}$ and quantile crossing series 
$\hat{u}_t(\al)/\sqrt{\al(1-\al)}$ instead.  The resulting spectrum takes the form $\tilde{S}(\om,\al) := S(\om,\al)/(\al(1-\al))$, which is more convenient for comparison across quantiles. For example,
when $\{ y_t \}$ is white noise, we have $S(\cdot,\al) = \al(1-\al)$, which depends on $\al$, 
but $\tilde{S}(\cdot,\al) = 1$, which is independent of $\al$.

The extensions of the SAR method to multiple time series and biquantile-crossing spectrum inherit the premise of offering more accurate spectral estimates when the underlying spectra are smooth in the quantile level.
Further analysis of these estimators and their applications is an interesting topic for future research.

The all-pole nature of the AR model determines that in practice  it is more effective for estimating 
certain types of spectral patterns (e.g., sharp peaks) and less so for other types (e.g., deep troughs).  
A more versatile extension of the AR model
for conventional spectral analysis is the ARMA model (Percival and Walden 1993; 
Stoica and Moese 1997).  How to leverage the ARMA model
to estimate the quantie-crossing spectrum as a two-dimensional function  
and overcome the restrictions of the SAR method 
is another interesting topic for future research. 

\newpage
\section*{Appendix I: On Assumption (A0)}

\setcounter{equation}{0}
\renewcommand{\theequation}{I.\arabic{equation}}

\noindent
{\it Proof of Theorem~\ref{thm:A0}}.
Because $\hat{u}_t(\al)$ is bounded in absolute value by 1, it suffices to show that
\eqn
n^{-1} \sum_{t=1}^n \hat{u}_t(\al) \hat{u}_{t-\tau}(\al) \plim R(\tau,\al)
\label{uu}
\eqqn
for fixed $\tau \in \{0,1,\dots\}$ and $\al \in (0,1)$.
Toward that end, we follow the approach of Linton and Whang (2007) and define
\eq
\hat{c}(x)  :=   n^{-1} \sum_{t=1}^n g(\bz_t,x), \quad c(x)  :=   \E\{ \hat{c}(x)  \}.
\eqq
Because $u_t(\al) = \psi_\al(y_t - q(\al))$, it follows that
\begin{gather*}
 n^{-1} \sum_{t=1}^n u_t(\al) \, u_{t-\tau}(\al)
 = n^{-1} \sum_{t=1}^n g(\bz_t,q(\al)) = \hat{c}(q(\al)), \\
  n^{-1} \sum_{t=1}^n \hat{u}_t(\al) \, \hat{u}_{t-\tau}(\al)
 = n^{-1} \sum_{t=1}^n g(\bz_t, \hat{q}(\al)) = \hat{c}(\hat{q}(\al)).
\end{gather*}
Moreover, $\E\{ g(\bz_t,q(\al))  \} = c(q(\al)) =  R(\tau,\al)$.
Therefore, under assumption (A2), we obtain
\eqn
\hat{c}(q(\al)) \plim R(\tau,\al).
\label{uu2}
\eqqn
The remaining objective is to show
\eqn
\hat{c}(\hat{q}(\al))-\hat{c}(q(\al)) = o_P(1).
\label{chat}
\eqqn
If this expression holds, we would have (\ref{uu}) as a result of (\ref{uu2}).

To prove (\ref{chat}), we observe that
\eq
c(x) & = & \E\{ (\al- I(y_t \le x)) \, (\al- I(y_{t-\tau} \le x)) \} \\
& = & \al^2 - \al \, \E\{ I(y_t \le x) \} - \al \, \E\{ I(y_{t-\tau} \le x) \} 
+ \E\{ I(y_t \le x, y_{t-\tau} \le x) \} \\
& = & \al^2 - 2\al F(x) + F_{\tau}(x,x).
\eqq
Under assumption (A4), we have
\eqn
\dot{c}(x) = - 2 \al \dot{F}(x) + \dot{F}_\tau(x,x).
\label{cdot}
\eqqn
Using the Taylor expansion of $c(x)$ at $\hat{q}(\al)$, we can write
\eq
c(q(a))  = c(\hat{q}(\al)) +
 \dot{c}( \tilde{q}(\al))\, (q(\al)-\hat{q}(\al)),
\eqq
where $\tilde{q}(\al)$ lies between $\hat{q}(\al)$ and $q(\al)$. This leads to
\eq
\hat{c}(\hat{q}(\al)) - c(q(\al)) 
= \hat{c}(\hat{q}(\al)) - c(\hat{q}(\al)) +
 \dot{c}( \tilde{q}(\al)) \, (\hat{q}(\al) - q(\al)).
\eqq
Therefore, 
\eqn
\hat{c}(\hat{q}(\al))-\hat{c}(q(\al)) & = &
[\hat{c}(\hat{q}(\al)) - c(q(\al)) ]
- [\hat{c}(q(\al)) - c(q(\al)) ] \notag \\
&=&
[\hat{c}(\hat{q}(\al)) - c(\hat{q}(\al))] +
 \dot{c}( \tilde{q}(\al)) \, (\hat{q}(\al) - q(\al))  \notag \\
 &&  - \ [\hat{c}(q(\al)) - c(q(\al)) ] \notag \\
 &:=& T_1 + T_2 - T_3.
 \label{T}
\eqqn
It  suffices to show that each of the three terms in (\ref{T}) takes the form $o_P(1)$.

Toward that end, we first note that under assumption (A2), 
\eqn
 \hat{c}(x) - c(x) = n^{-1} \sum_{t=1}^n [g(\bz_t,x) - \E\{g(\bz_t,x)\}] = o_P(1)
\label{lln}
\eqqn
for any fixed $x \in \bbR$. This result implies
\eqn
T_3 := \hat{c}(q(\al)) - c(q(\al))   = o_P(1).
\label{T3}
\eqqn
Moreover, combining (\ref{cdot}) with assumptions (A1) and (A4) gives
\eq
\dot{c}(\tilde{q}(\al)) \plim \dot{c}(q(\al)) =  - 2\al \dot{F}(q(\al)) 
+ \dot{F}_\tau(q(a),q(a)). 
\eqq
This, together with assumption (A1), yields
\eqn
 T_2 := \dot{c}( \tilde{q}(\al)) \, (\hat{q}(\al) - q(\al))  = o_P(1).
 \label{T2}
\eqqn
Finally, if (\ref{lln}) could be strengthened to uniform convergence, i.e., if
\eqn
 \sup_{x \in \Theta} |\hat{c}(x)-c(x)| \plim 0
\label{uniform}
\eqqn
for any compact set $\Theta \subset \bbR$. Then, for any $\del > 0$, we would have
\eq
\Pr\{ |\hat{c}(\hat{q}(\al))-c(\hat{q}(\al)) | > \del \} 
& = & \Pr\{ |\hat{c}(\hat{q}(\al))-c(\hat{q}(\al)) |  > \del, |\hat{q}(\al) - q(\al)| > \del \} \\
& & + \ \Pr\{ |\hat{c}(\hat{q}(\al))-c(\hat{q}(\al)) |  > \del, |\hat{q}(\al) - q(\al)| \le \del \} \\
& \le & \Pr\{ |\hat{q}(\al) - q(\al)| > \del \} 
+ \Pr\bigg\{ \sup_{x \in \Theta} |\hat{c}(x)-c(x) | > \del \bigg\},
\eqq
where $\Theta := \{x: |x-q(\al)| \le \del\}$. Combining this with 
 (A1) and (\ref{uniform}) would yield
\eq
\lim_{n\linfty} \Pr\{  |\hat{c}(\hat{q}(\al))-c(\hat{q}(\al)) |  > \del \} = 0
\eqq
for any $\del > 0$, which  means, by definition,
\eqn
T_1 :=  \hat{c}(\hat{q}(\al)) - c(\hat{q}(\al)) = o_P(1).
\label{T1}
\eqqn
The assertion (\ref{chat}) would follow immediately upon collecting (\ref{T1}), (\ref{T2}), and (\ref{T3}).
Therefore, the remaining task is to establish the uniform convergence  (\ref{uniform}).

To prove the uniform convergence, we resort to a stochastic equicontinuity argument (P\"{o}tscher and Prucha 1994).
First, for fixed $\bz \in Z_1 := \{(z_1,z_2): z_1 < z_2\}$,  we can write
\eq
g(\bz,x) = \left\{
\begin{array}{ll}
\al^2 & \text{ if $x \in D_1(\bz)$}, \\
\al (\al-1) & \text{ if $x\in D_2(\bz)$}, \\
(\al-1)^2 & \text{ if $x\in D_3(\bz)$}.
\end{array} \right.
\eqq
where $D_1(\bz) := (-\infty,z_1)$, 
$D_2(\bz) := [z_1,z_2)$, and $D_3(\bz) := [z_2,\infty)$. 
This implies
\eq
d(\bz,x,x') := g(\bz, x) - g(\bz, x') 
 = \left\{
\begin{array}{ll}
0 & \text{ if  $x \in D_1(\bz)$ and $x' \in D_1(\bz)$}, \\
\al   &  \text{ if $x \in D_1(\bz)$ and $x' \in D_2(\bz)$}, \\
2\al-1   &  \text{ if $x \in D_1(\bz)$ and $x' \in D_3(\bz)$}, \\
-\al & \text{ if  $x \in D_2(\bz)$ and $x' \in D_1(\bz)$}, \\
0  &  \text{ if $x \in D_2(\bz)$ and $x' \in D_2(\bz)$}, \\
\al-1   &  \text{ if $x \in D_2(\bz)$ and $x' \in D_3(\bz)$}, \\
1-2\al & \text{ if  $x \in D_3(\bz)$ and $x' \in D_1(\bz)$}, \\
1-\al  &  \text{ if $x \in D_3(\bz)$ and $x' \in D_2(\bz)$}, \\
0   &  \text{ if $x \in D_3(\bz)$ and $x' \in D_3(\bz)$}.
\end{array} \right.
\eqq
For fixed $x \in D_1(\bz)$, we can write
\eq
\sup_{x' \in B(x,\del)} |d(\bz, x, x')| = 
\left\{
\begin{array}{ll}
0 & \text{ if $D_2(\bz) \cap B(x,\del) = \emptyset$ and 
$D_3(\bz) \cap B(x,\del) = \emptyset$,} \\
\al & \text{ if $D_2(\bz) \cap B(x,\del) \ne \emptyset$ and $D_3(\bz) \cap B(x,\del) = \emptyset$,} \\
|2\al-1| & \text{ if $D_2(\bz) \cap B(x,\del) =\emptyset$ and $D_3(\bz) \cap B(x,\del) \ne \emptyset$,} \\
\max\{ \al,|2\al-1|\} & \text{ if $D_2(\bz) \cap B(x,\del) \ne \emptyset$ and $D_3(\bz) \cap B(x,\del) \ne \emptyset$.} 
\end{array} \right.
\eqq
The nonzero values in this expression are always smaller than 1, and they are taken if only if $D_2(\bz)  \cup D_3(\bz)) \cap B(x,\del)\ne \emptyset$. Therefore,
\eq
\sup_{x' \in B(x,\del)} |d(\bz, x, x')| \le I( (D_2(\bz)  \cup D_3(\bz)) \cap B(x,\del)) 
= I(x < z_1 \le x+\del).
\eqq
Similarly, for  $x \in D_2(\bz)$, we have
\eq
\sup_{x' \in B(x,\del)} |d(\bz, x, x')| & \le &
 I( (D_1(\bz) \cup D_3(\bz)) \cap B(x,\del)) \\
 & = & I(x-\del < z_1 \le x < z_2 \le x+\del) \\
 & \le & I( x < z_2 \le x+\del),
\eqq
and for  $x \in D_3(\bz)$, we have 
\eq
\sup_{x' \in B(x,\del)} |d(\bz, x, x')| \le  I( (D_1(\bz) \cup D_2(\bz)) \cap B(x,\del))
= I(x-\del < z_2 \le x).
\eqq
Therefore, for any given $x \in \Theta$, we can write
\eq
\sup_{x' \in B(x,\del)} |d(\bz, x, x')|
\le I(x < z_1 \le x+\del) +  I( x < z_2 \le x+\del)  + I( x -\del < z_2 \le x).
\eqq
This implies
\eq
\lefteqn{
\E\bigg\{ 
\sup_{x' \in B(x,\del)} |d(\bz_t, x, x')| I(\bz_t \in Z_1) \bigg\}} \\
&  \le & \E\{ [I(x < y_t \le x+\del) +  I( x < y_{t-\tau} \le x+\del)  + I( x-\del < y_{t-\tau} \le x)]
I(\bz_t \in Z_1) \} \\
& \le & \E\{ I(x < y_t \le x+\del) +  I( x < y_{t-\tau} \le x+\del)  + I( x-\del < y_{t-\tau} \le x) \} \\
& = & 2(F(x+\del)-F(x)) + (F(x)-F(x-\del)) \\
& \le & \kappa \del,
\eqq
where $\kappa := 3 \max\{ \dot{F}(x): x\in \bbR\}$. 
Due to symmetry, the same result can be obtained when $\bz_t$ is restricted to $ Z_2 := \{ (z_1,z_2): z_2 < z_1\}$.
Combining these results yields
\eq
\E \bigg\{\sup_{x' \in B(x,\del)} |d(\bz_t, x, x')| \bigg\} 
& = &
\E \bigg\{\sup_{x' \in B(x,\del)} |d(\bz_t, x, x')|  \, I(\bz_t \in Z_1)\bigg\} \notag \\
& & +\ \E \bigg\{\sup_{x' \in B(x,\del)} |d(\bz_t, x, x')| \, I(\bz_t \in Z_2)  \bigg\} \notag \\
& \le & 2 \kappa \del.
\eqq
This, together with the stationarity of $d(\bz_t, x, x')$, implies
\eqn
\sup_{x\in \Theta} \bigg( \limsup_{n\linfty} \,
n^{-1} \sum_{t=1}^n \E \bigg\{\sup_{x' \in B(x,\del)} |d(\bz_t, x, x')| \bigg\}  \bigg)
\le 2\kappa \del \lzero
\label{ec3}
\eqqn
as $\del \lzero$. 
Combining this with assumption (A3), which is the condition (4.7) in P\"{o}tscher and Prucha (1994) (see also Andrews (1987)), guarantees the uniform convergence (\ref{uniform}) according to P\"{o}tscher and Prucha (1994, Corollary 4.4). \qed

\begin{rem}
A typical approach to proving  a uniform convergence result such as (\ref{uniform}) is through a more stringent  stochastic equicontinuity condition of the form
\eqn
\lim_{\del \lzero}
\limsup_{n \linfty} \, n^{-1} \sum_{t=1}^n
\Pr \bigg\{ \sup_{x\in \Theta} \, \sup_{x' \in B(x,\del)} | g(\bz_t,x') - g(\bz_t,x) | > \ep \bigg\} = 0
\label{uec2}
\eqqn
or 
\eqn
\lim_{\del \lzero}
\limsup_{n \linfty} \, n^{-1} \sum_{t=1}^n
\E \bigg\{ \sup_{x\in \Theta} \, \sup_{x' \in B(x,\del)} | g(\bz_t,x') - g(\bz_t,x) | \bigg\} = 0,
\label{uec}
\eqqn
These conditions are more difficult to verify than (\ref{ec3}). One often resorts to a  Lipschitz condition (Newey 1991; P\"{o}tscher and Prucha 1994). A number of ``primitive'' conditions are provided in 
Andrews (1992) that imply (\ref{uec2}). All these conditions require $g(\bz,x)$ to be uniformly continuous in $x \in \Theta$. Unfortunately, this requirement  is not fulfilled  by $g(\bz,x)$ in (\ref{g}). In fact, given $\bz := (z_1,z_2)$, 
the function $g(\bz,x)$ is piecewise constant in $x  \in \bbR$ with jumps at $x=z_1$ and $x=z_2$. 
\end{rem}

\begin{rem}
Under a stochastic volatility model,  an assertion was made by Linton and Whang (2007, Eq.\ 14) that would 
imply (\ref{T1}) when combined with (\ref{T3}). Unfortunately, no details were provided in their proof to justify this assertion. 
In our proof of (\ref{T3}), the less-stringent
stochastic equicontinuity condition (\ref{ec3})  is verified, which leads to the uniform convergence result (\ref{uniform}) when combined with assumption (A3). 
\end{rem}

\bigskip
Assumptions (A2) and (A3) are satisfied by stationary processes with the strong mixing 
property when the serial dependence vanishes sufficiently fast.

\begin{pro}
Assumptions (A2) and (A3) are satisfied if $\{ y_t \}$ is a stationary process with the strong mixing property 
(\ref{mixing}) such that $\sum_{s=1}^\infty m_s < \infty$.
\label{pro:A2A3}
\end{pro}

\noindent
{\it Proof}.   If $\{ y_t \}$ is strong mixing
with mixing numbers $m_s$ $(s=1,2,\dots)$, then $\{ \bz_t \}$, defined by (\ref{z}), is also strong mixing, and its mixing number at lag $s$ for any $s > \tau$
is less than or equal to $m_{s-\tau}$ because $\vsig(\bz_t: t \le 0) \subseteq \vsig(y_t: t \le 0)$
and $\vsig(\bz_t: t \ge s) \subseteq \vsig(y_t: t \ge s -\tau)$.
For any measurable function $h(\cdot)$, $h(\bz_t)$  is also strong 
mixing, and its mixing number at lag $s$ is less than or equal to the mixing number 
of $\bz_t$ at lag $s$  because
$\vsig(h(\bz_t): t \le 0) \subseteq \vsig(\bz_t: t \le 0)$ and $\vsig(h(\bz_t): t \ge s) \subseteq \vsig(\bz_t: t \ge s)$.
The latter is bounded from above by $m_{s-\tau}$ for any $s > \tau$. Furthermore,
if $h(\cdot)$ is bounded, i.e., if there exists a constant $h_0  > 0$ such that $|h(\cdot)| \le h_0$, 
then, an application of the mixing inequality (Billingsley 1995, p.\ 365) gives
\eqn
|\Cov\{h(\bz_t),h(\bz_{t-s})\}| \le 4 \, h_0^2 \, m_{s-\tau} \quad (s > \tau).
\label{cov}
\eqqn
This implies that the ACF of $\{ h(\bz_t) \}$ is absolutely summable if $\sum_{s=1}^\infty m_s < \infty$.
Under this condition, the weak law of large numbers for stationary processes
(Li 2014, Lemma 12.5.3, p.\ 589) can be cited to obtain
\eq
n^{-1} \sum_{t=1}^n [ h(\bz_t) - \E\{ h(\bz_t) \}] \plim 0
\eqq
as $n \linfty$. The assertion follows by considering the following choices of $h(\cdot)$: (a) $h(\bz) := g(\bz,x) := \psi_\al(z_1-x)  \psi_\al(z_2-x)$ with $\bz := (z_1,z_2) \in \bbR^2$ for fixed $x \in \bbR$ and $\al \in (0,1)$, where $\psi_\al(x) := \al - I(x \le 0)$; (b) $h(\bz) := g_1(\bz,x,\del) := \inf\{ |g(\bz,x')|: x' \in B(x,\del)\}$, where
$B(x,\del) := \{x' \in \Theta: |x'-x| \le \del \}$; 
(c) $h(\bz) := g_2(\bz,x,\del) := \sup\{ |g(\bz,x')|: x' \in B(x,\del)\}$. \qed

\bigskip
The strong mixing property in Proposition~\ref{pro:A2A3}, together with assumption (A4) regarding $F(\cdot)$, 
is sufficient to guarantee assumption (A1).

\begin{pro}
Assumption (A1) is satisfied if (a) $\{ y_t \}$ is a stationary process with the strong mixing property (\ref{mixing}) such that $\sum_{s=1}^\infty  m_s < \infty$
and (b)  $F(\cdot)$ satisfies  assumption (A4).
\label{pro:A1}
\end{pro}

\noindent
{\it Proof}. Define $\tilde{\psi}_\al(x) := (1-\al) I(x < 0) - \al I(x >0)$. 
Let $\Psi_n(q) := n^{-1} \sum_{t=1}^n \tilde{\psi}_\al(y_t - q)$ for $q \in \bbR$.
The stationarity of $\{ y_t \}$ and the continuity of $F(\cdot)$ imply that 
 $\Psi(q) := \E\{ \Psi_n(q) \} = \E\{ \tilde{\psi}_\al(y_t-q) \} = F(q) - \al$ for all $q \in \bbR$.
By definition, $\Psi_n(\hat{q}(\al) )= o_P(1)$ (van der Vaart 1998, p.\ 43) and 
$\Psi(q(\al) ) = 0$. Moreover, for any fixed $q$, condition (a) implies that $\{ \tilde{\psi}_\al(y_t - q) \}$ is a stationary 
process with an absolutely summable autocovariance function. Therefore,  by the weak law of large numbers for stationary processes (Li 2014, Lemma 12.5.3, p.\ 589), we have $\Psi_n(q) \plim \Psi(q)$ 
for any fixed $q$. In addition, under condition (b), we have $\Psi(q(\al)-\ep) < \Psi(q(\al) ) = 0 < \Psi(q(\al)+\ep)$
for any  $\ep > 0$.   Combining these results with the fact that $\Psi_n(q)$ as a nondecreasing 
function of $q$ leads to the assertion, citing  the consistency theorem of $M$-estimators 
(van der Vaart 1998, Lemma 5.10, p.\ 47). \qed

\section*{Appendix II: R Functions}

All spectral estimators discussed in this paper are implemented 
in the R package `qfa' (version $\ge$ 4.1) which can be installed from \url{https://cran.r-project.org} (also available
at \url{https://github.com/thl2019/QFA}).
Relevant functions are described below. They are applicable to univariate as well as multivariate time series.
\begin{itemize}
\item {\tt qcser}: a function that takes a time series as input and creates its quantile-crossing series (QCSER) at a given set of quantile levels.
\item {\tt qser2qacf}: a function that takes the QCSER from {\tt qcser} as input and produces the corresponding 
sample autocovariance functions (ACFs).
\item {\tt qser2sar}: a function that takes the QCSER as input and produces the SAR estimates 
of AR parameters  at a given set of quantile levels for construction of spectral estimates. 
\item {\tt qspec.lw}: a function that takes the ACFs from {\tt qer2qacf} as input and produces the lag-window (LW) estimate of the quantile-crossing spectrum evaluated at the given quantile levels and all Fourier frequencies. With the option {\tt M=NULL}, this function produces the ordinary periodogram of the QCSER.
\item {\tt qspec.ar}: a function  that takes the QCSER from {\tt qcser} as input and produces the AR estimate of the quantile-crossing spectrum  evaluated at the given quantile levels and all Fourier frequencies. With the option
{\tt method="sp"}, it produces the AR-S estimate by post-smoothing the AR parameters using the R function {\tt smooth.spline}. With the option {\tt p=NULL}, the order parameter $p$ is automatically selected by AIC. 
\item {\tt qspec.sar}: a function that takes the QCSER from {\tt qcser} as input and produces the SAR estimate of the quantile-crossing spectrum  evaluated at the given quantile levels and all Fourier frequencies. With the option {\tt p=NULL}, the order parameter $p$ is automatically selected by AIC. In this implementation, the integral in the penalty term is replaced by the sum over the quantile levels.
\item {\tt qfa.plot}: a function that takes a quantile-crossing spectrum or its estimate as input and displays it as an image, with horizontal axis representing the frequency and vertical axis representing the quantile level.
\end{itemize}

\section*{References}

{\footnotesize
\begin{description} 

\item
Akaike H. (1969). Power spectral estimation through autoregressive model fitting. 
{\it Annals of the Institute of Statistical Mathematics}, 21, 407--419.

\item 
Andrews, D. (1987). Consistency in nonlinear econometric models: A generic uniform law of
large numbers. {\it Econometrica}, 55, 1465--1471.

\item
Andrews, D. (1992). Generic uniform convergence. {\it Econometric Theory}, 8, 241--257.

\item
Athreya, K., and Pantula, S. (1986). A note on strong mixing of ARMA processes. 
{\it Statistics \& Probability Letters}, 4, 187--190.

\item
Bantli, F., and Hallin, M. (1999). $L_1$-estimation in linear models with heterogeneous white noise,
{\it Statistics \& Probability Letters}, 45, 305--315.

\item
Berk, K. (1974). Consistent autoregressive spectral estimates. {\it Annals of  Statistics}, 2, 489--502.

\item
Billingsley, P. (1994). {\it Probability and Measure} (Third Edition). New York: Wiley.

\item
Birr, S., Volgushev, S., Kley, T., Dette, H., and Hallin, M. (2017). Quantile spectral
analysis for locally stationary time series. 
{\it Journal of Royal Statistical Society Series B}, 79, 1619--1643.

\item
Brillinger, D. (1968). Estimation of the cross-spectrum of a stationary bivariate Gaussian process from its zeros.
{\it Journal of the Royal Statistical Society Series B}, 30, 145--159.

\item 
Brockwell, P., and Davis, R. (1991). {\it Time Series: Theory and Methods} (Second Edition). New York: Springer.

\item
Chen, T., Sun, Y., and Li, T.-H. (2021). A Semi-parametric estimation method for the quantile
spectrum with an application to earthquake classification  using convolutional neural network.
{\it Computational Statistics \& Data Analysis}, 154, 107069.

\item
Choi, B. (1993). Multivariate maximum entropy spectrum. {\it Journal of Multivariate Analysis}, 46, 56--60.

\item
Davis, R., and Mikosch, T. (2009). The extremogram: A correlogram for extreme events. 
{\it Bernoulli}, 15, 977--1009.

\item 
Dette, H., Hallin, M., Kley, T., and Volgushev, S. (2015). Of copulas,
quantiles, ranks and spectra: an $L_1$-approach to spectral analysis. 
{\it Bernoulli}, 21, 781--831.

\item
Goto, Y., Kley, T., Van Hecke, R., Volgushev, S., Dette, H., and Hallin, M. (2022). The
integrated copula spectrum. {\it Annals of  Statistics}, 50, 3563--3591.

\item
Hagemann, A. (2013). Robust spectral analysis. arXiv:1111.1965.

\item
Hastie, T., and Tibshirani, R. (1990). {\it Generalized Additive Models}. London:  Chapman \& Hall.

\item
Hinich, M. (1967). Estimation of spectra after hard clipping of Gaussian processes. 
{\it Technometrics},  9, 391--400.

\item
Hong, Y. (2000). Generalized spectral tests for serial dependence. 
{\it Journal of the Royal Statistical Society Series B}, 62, 557--574. 

\item
Jim\'{e}nez-Var\'{o}n, C., Sun, Y., and Li, T.-H. (2024).
A semi-parametric estimation method for  quantile coherence  with an application to bivariate financial time series clustering. 
{\it Econometrics and Statistics}. \url{https://doi.org/10.1016/j.ecosta.2024.11.002}

\item
Jordanger, L., and Tj{\o}stheim, D. (2022). Nonlinear spectral analysis: a local
Gaussian approach. {\it  Journal of the American Statistical Association}, 117, 1010--1027. 

\item
Jordanger, L., and Tj{\o}stheim, D. (2023). Local Gaussian cross-spectrum analysis.
{\it Econometrics}, 11, 1--27.

\item
Kedem, B. (1986). Spectral analysis and discrimination by zero-crossings. {\it Proceedings of the IEEE},  74, 1477--1493.

\item
Kley, T., Volgushev, S., Dette, H., and Hallin, M. (2016). Quantile spectral processes:
Asymptotic analysis and inference. {\it Bernoulli}, 22, 2016, 1770--1807.

\item
Koenker, R. (2005).  {\it Quantile Regression}.  Cambridge, UK: Cambridge University Press.

\item
Li, T.-H. (2008). Laplace periodogram for time series analysis.
 {\it Journal of the American Statistical Association}, 
103, 757--768.

\item
Li, T.-H. (2012). Quantile periodograms. 
{\it Journal of the American Statistical Association}, 107, 765--776.

\item
Li, T.-H. (2014). {\it Time Series with Mixed Spectra}. Boca Raton, FL: CRC Press.

\item 
Li, T.-H. (2020). From zero crossings to quantile-frequency analysis of time
series with an application to nondestructive evaluation. 
{\it Applied Stochastic Models for Business and Industry}, 36, 1111--1130.

\item
Li, T.-H. (2025a). Quantile Fourier transform, quantile series, and nonparametric estimation of quantile spectra.
{\it Communications in Statistics - Simulation and Computation}, DOI: 10.1080/03610918.2025.2509820

\item
Li, T.-H. (2025b). Spline autoregression method for estimation of
quantile spectrum. {\it Journal of Computational and Graphical Statistics}, DOI: 10.1080/10618600.2025.2549452

\item
Linton, O., and Whang, Y.-J. (2007). The quantilogram: With an application to evaluating directional predictability.
{\it Journal of Econometrics}, 141, 250--282.

\item
Mokkadem, A. (1988). Mixing properties of ARMA processes. 
{\it Stochastic Processes and their Applications}, 29, 309--315.

\item
Newey, W. (1991). Uniform convergence in probability and stochastic equicontinuity.
{\it Econometrica}, 59, 1161--1167.

\item
Parzen, E. (1969). Multiple time series modeling. In {\it Multivariate Analysis II} (P. Krishnaiah, ed.). New York: Academic Press.

\item
Parzen, E. (1982). Maximum entropy interpretation of autoregressive spectral densities. 
{\it Statistics \& Probability Letters}, 1, 7--11.

\item
Percival, D., and Walden, A. (1993). {\it Spectral Analysis for Physical Applications}, Chapter 9.
Cambridge, UK: Cambridge University Press.

\item
P\"{o}tscher, B., and Prucha, I. (1994). Generic uniform convergence and equicontinuity concepts for random functions.
{\it Journal of Econometrics}, 60, 23--63.

\item
Priestley, M. (1981). {\it Spectral Analysis and Time Series}. New York: Academic Press. 

\item
R Core Team (2024). R: A language and environment for statistical computing. R Foundation for Statistical Computing, Vienna, Austria.  URL https://www.R-project.org/.

\item
Rosenblatt, M. (1956). A central limit theorem and a strong mixing condition. {\it Proceedings
of the National Academy of Sciences of the United States of America}, 42, 43--47.

\item
Stoica, P., and Moses, R. (1997). 
{\it Introduction to Spectral Analysis}, Chapter 3. Upper Saddle River, NJ: Prentice Hall.

\item
van der Vaart, A. (1998). {\it Asymptotic Statistics}. Cambridge, UK: Cambridge University Press.

\item
Wahba, G. (1990) {\it  Spline Models for Observational Data}. SIAM. CBMS-NSF Regional Conference
Series in Applied Mathematics, v.\ 59.

\item
Withers, C. (1981). Conditions for linear processes to be strong-mixing. 
{\it Zeitschrift f\"{u}r Wahrscheinlichkeitstheorie und verwandte Gebiete}, 57, 477--480.

\end{description}
}

\end{document}